\documentclass[%
 reprint,
 amsmath,amssymb,
 aps,
]{revtex4-2}

\usepackage{graphicx}
\usepackage{dcolumn}
\usepackage{bm}

\usepackage{amsmath}
\usepackage{mathtools}
\usepackage{cases}
\usepackage{caption}
\usepackage{comment}
\usepackage{array}
\usepackage{tabularx}
\usepackage{makecell}
\usepackage{mathrsfs}
\usepackage{parskip}
\usepackage{gensymb}
\usepackage[separate-uncertainty=true,separate-uncertainty-units=single]{siunitx}
\DeclareSIUnit{\wtpercent}{wt\%}

\usepackage[hidelinks]{hyperref}
\usepackage[noabbrev]{cleveref}
\usepackage{xr}
\def\b{\bm}

\def\Er{\mbox{Er}}

\def\f{\b{f}}

\def\h{\b{h}}

\def\m{\b{m}}
\def\n{\b{n}}

\def\v{\b{v}}
\def\x{\b{x}}

\def\F{\b{F}}

\def\V{\b{V}}
\def\X{\b{X}}

\def\I{\b{\mathrm{I}}}

\def\Pm{\b{\mathrm{P}}}
\def\P{\Pm}

\def\Tm{\b{\mathrm{T}}}

\def\bbeta{\bm{\hat{\eta}}}

\def\bnu{\bm{\hat{\nu}}}
\def\bPi{\bm{\Pi}}

\def\nhat{\b{\hat{n}}}
\def\rhat{\hat{\b{r}}}
\def\shat{\b{\hat{s}}}
\def\xhat{\hat{\b{x}}}
\def\yhat{\hat{\b{y}}}
\def\zhat{\hat{\b{z}}}

\def\xihat{\hat{\b{\xi}}}
\def\zehat{\hat{\b{\zeta}}}

\def\de{\mathrm{d}}

\def\dd{\mathrm{d}}

\def\Oh{\mathcal{O}}
\def\im{i}

\usepackage{color}

\begin{document}

\preprint{APS/123-QED}

\title{Morphogenesis of bacterial colonies in liquid crystalline environments}

\author{Sebastian Gonzalez La Corte}
\email{These authors contributed equally to this work.}
\affiliation{Lewis-Sigler Institute for Integrative Genomics \\
 Princeton University, Princeton, NJ, USA}

\author{Thomas G.\ J.\ Chandler}
\email{tgjchandler@unc.edu}
\email{These authors contributed equally to this work.}
\affiliation{Department of Mathematics, \\
University of North Carolina at Chapel Hill, Chapel Hill, NC, USA} 


\author{Saverio E.\ Spagnolie}

\email{spagnolie@math.wisc.edu}
\affiliation{Department of Mathematics \\
University of Wisconsin--Madison, Madison, WI, USA} 

\author{Ned S.\ Wingreen}
\email{wingreen@princeton.edu}
\affiliation{
Department of Molecular Biology and\\ 
Lewis-Sigler Institute for Integrative Genomics,\\ 
Princeton University, Princeton, NJ, USA
}%

\author{Sujit S.\ Datta}
\email{ssdatta@caltech.edu}
\affiliation{%
Division of Chemistry and Chemical Engineering \\ California Institute of Technology, Pasadena, CA, USA
}%
\affiliation{%
Department of Chemical and Biological Engineering \\ Princeton University, Princeton, NJ, USA
}%

\date{\today}

\begin{abstract}
Natural bacterial habitats are often complex fluids with viscoelastic and anisotropic responses to stress; for example, they can take the form of liquid crystals (LCs), with elongated microscopic constituents that collectively align while still retaining the ability to flow. However, laboratory studies typically focus on cells in simple liquids or complex fluids with randomly-oriented constituents.
Here, we show how interactions with
LCs shape bacterial proliferation in multicellular colonies.
Using experiments, we find that in a
nematic LC, cells generically form aligned single-cell-wide ``chains” as they reproduce. As these chains lengthen, they eventually buckle in a highly localized manner. By combining our measurements with a continuum mechanical theory, we demonstrate that this distinctive morphogenetic program emerges because cells are kept in alignment due to the LC's elasticity; as each chain lengthens, growth-induced viscous stresses along its contour eventually overcome the elasticity of the surrounding nematic, leading to buckling. 
Our work thus reveals and provides mechanistic insight into the previously-overlooked role of LCs in sculpting bacterial life in complex environments.
\end{abstract}

\maketitle


\section{Introduction}
Many bacterial habitats can exhibit liquid crystalline (LC) order; prominent examples include host mucus~\cite{figueroa2019mechanical,viney1993liquid,davies1998water}, extracellular polymeric substances in biofilms~\cite{repula2022biotropic}, and dense suspensions of filamentous phages~\cite{secor2015biofilm,secor2015filamentous,tarafder_phage_2020}. However, laboratory studies of bacteria typically focus on cells in isotropic environments.
As a result, whether and how
inhabiting a liquid crystal influences bacterial behavior
remains poorly understood.

Recent work indicates that the anisotropic rheology of LCs can dramatically alter how individual cells swim~\cite{mushenheim2014dynamic,zhou2014living,kst15,genkin2017topological,lcg21,goral2022frustrated,prabhune2024bacteria,mou2025simulations} and aggregate with other cells~\cite{mushenheim2014dynamic,gsla17,turiv2020polar,cgba22,fgsa22,los25,baza2025bend,wu2025programmable}. 
Nevertheless, despite growing recognition that an environment's mechanical properties, in addition to its chemistry, can influence bacterial growth~\cite{Fung2013,koch2013bacterial,fraej07,ypsg18,hdddeww23,wplkfs24}, the possible influence of liquid crystalline order on another fundamental characteristic of bacterial life---proliferation through space in a multicellular colony~\cite{shapiro1988bacteria}---remains underexplored. 
Addressing this gap in knowledge is important to establish mechanistic principles governing how proliferating active matter interacts with complex environments
~\cite{hallatschek2023proliferating,su23}. It also has key biological implications; indeed, in nature, many bacteria are nonmotile or lose motility~\cite{folkesson2012adaptation,rau2010early,cullen2015bacterial,blair2008molecular,sauer2002pseudomonas}, but continue to proliferate in colonies. Colony morphology in turn influences how cells access nutrients, cooperate and compete with each other, resist external stressors (e.g., antibiotics, predators, phage), and cause infections in hosts~\cite{shapiro1988bacteria,fei_nonuniform_2020,trejo_elasticity_2013,martinez-calvo_morphological_2022,fujikawa_fractal_1989,farrell_mechanically_2013,wplkfs24,bottura_intra-colony_2022, hallatschek_genetic_2007,muller_genetic_2014,farrell_mechanical_2017,secor_entropically_2018,whiteley_gene_2001,mah_genetic_2003,Young06,Young07,slk90,hh99,psvpssp99,cj06}. Hence, we ask: {Can an environment's liquid crystalline order} affect the morphology of proliferating bacterial colonies?

\begin{figure*}
\includegraphics[width=\textwidth]{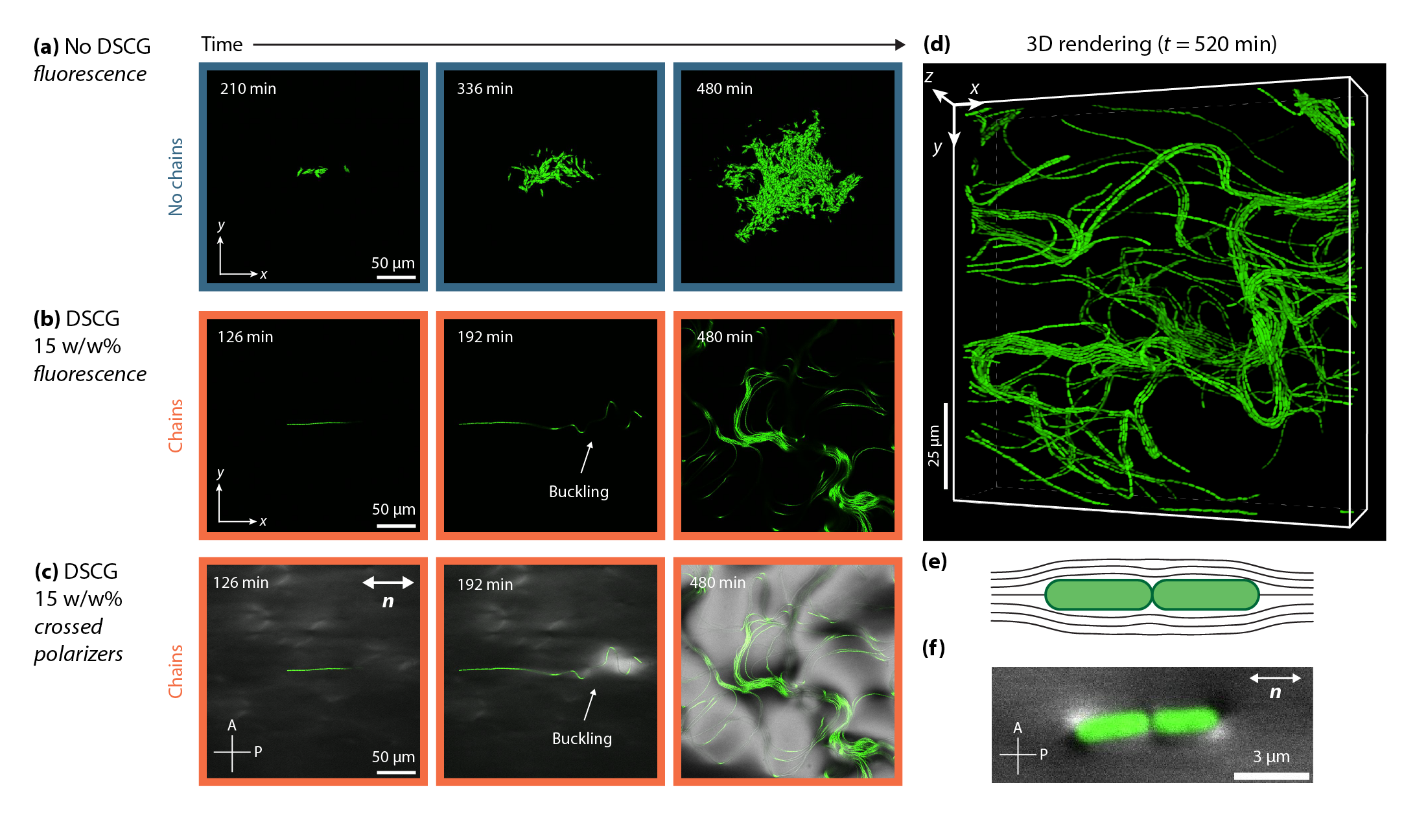}
\caption{\textbf{Proliferating nonmotile bacteria form chain-like colonies in liquid crystalline environments.} \textbf{(a)} Fluorescent confocal microscopy time sequence of nonmotile \textit{E.~coli} proliferating in nutrient-rich LC-free fluid; cells form an isotropic dispersion as they grow, divide, and diffuse apart. \textbf{(b)}  Same as in \textbf{a}, but in nematic DSCG solution (\SI{15}{w/w\%}); cells instead remain aligned end-to-end, forming a single-cell-wide chain that eventually buckles. \textbf{(c)} Same as in \textbf{b} with crossed-polarizer imaging overlaid (directions indicated by A and P), revealing local reorientation of the LC director field $\boldsymbol{n}$ accompanying chain buckling (bright regions). The overall far-field orientation imposed on the director by the underlying substrate is shown by the double-headed arrow. \textbf{(d)}~Three-dimensional confocal reconstruction of a chain-like colony after \SI{520}{\min}, showing the serpentine, single-cell-wide internal structure. \textbf{(e)} Schematic of the LC director profile around a rod-shaped bacterium with weak tangential anchoring; boojums (topological defects) form at the cell poles where the director field must accommodate the curved cell surface. \textbf{(f)} Fluorescent confocal micrograph with crossed-polarizer imaging overlaid (directions indicated by A and P) of two cells dispersed in nematic DSCG solution. The image shows the local reorientation of the LC director field $\boldsymbol{n}$ around the cell, with the emergence of boojums at the cell poles. The overall far-field orientation imposed on the director by the underlying substrate is shown by the double-headed arrow. Experiments showed are performed using a microfluidic device \SI{25}{\um } in height, \SI{25}{\mm} in width, and \SI{22}{\mm} in length.
}
\label{fig:fig_expt_chain}
\end{figure*}

Here, we demonstrate that this is indeed the case, and we elucidate the underlying biophysical mechanisms. Through experiments, we find that rod-shaped nonmotile bacteria form single-cell-wide ``chains'' as they proliferate in a nematic LC, all aligned along the LC director---in stark contrast to forming a random dispersion as in the conventionally studied non-LC case. The chains elongate as  proliferation continues, eventually buckling after reaching a critical length. This mechanical instability occurs over a much smaller length scale within the chain than the typical Euler buckling exhibited by long, slender filaments under compression. Using a continuum mechanical theory, we trace the origin of chain formation and buckling to the competition between LC elasticity, which tends to keep cells aligned along the LC director, and growth-induced compression from frictional stresses along each lengthening chain. This work thereby reveals how liquid crystalline environments can dramatically shape proliferating bacterial colonies and provides quantitative principles to predict and control these morphodynamics more broadly.

\section{Experimental Results}


\begin{figure*}
\includegraphics[width=0.8\textwidth]{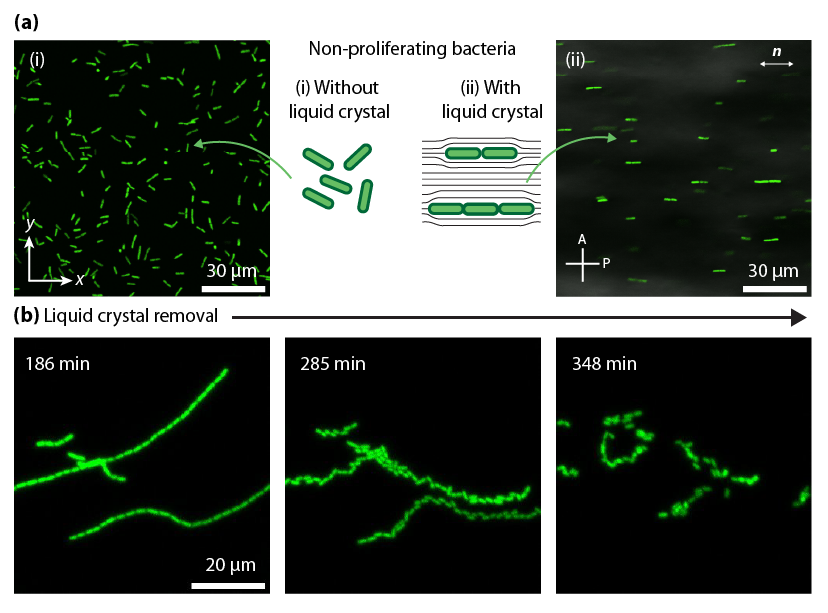}
\caption{\textbf{LC-mediated elastic forces are required for chain formation.} \textbf{(a)}  Non-proliferating \textit{E.~coli} cells (i)~remain randomly dispersed in LC-free fluid but (ii) spontaneously align and form end-to-end aggregates in nematic DSCG (\SI{18}{w/w\%}). Images show confocal micrographs with crossed-polarizer imaging overlaid in (ii) as in Fig.~\ref{fig:fig_expt_chain}(c),(f). \textbf{(b)} Time sequence showing chain disintegration after LC removal; LC-free solvent is introduced at $t = 0$. Cells remain in smaller side-by-side aggregates after LC removal, likely due to residual surface adhesin interactions~\cite{chekli2023escherichia,nwoko2021bacteria} that persist after cells are brought together by LC-induced elastic forces. See Movie~S\ref{vid:dilution video} for full field of view and dynamics.}
\label{fig:fig_lc_forces} 
\end{figure*}

\textbf{Nonmotile bacteria form long, single-cell-wide chains as they proliferate in a nematic liquid crystal.} Our experiments probe the proliferation of nonmotile bacteria in nutrient-rich aqueous solutions housed in thin, flat polydimethylsiloxane (PDMS) imaging channels bonded to underlying glass substrata. To start, we use \emph{Escherichia coli} as a model bacterium; the cells constitutively express green fluorescent protein (GFP) in their cytoplasm, enabling direct visualization using fluorescence confocal microscopy. In LC-free solution, the cells continually grow, divide, separate from each other, and diffuse apart, eventually forming an isotropic dispersion [Fig.~\ref{fig:fig_expt_chain}(a), Movie~S\ref{vid:lb video growth}]. 

We observe dramatically different behavior upon addition of disodium chromoglycate (DSCG), a nontoxic lyotropic LC~\cite{cheng2005compatibility,helfinstine2006lyotropic}, to the solution at a concentration $c=\SI{15}{w/w\%}$. Under these conditions, the solution forms a nematic~\cite{zimmermann2015self}, in which the chromonic DSCG molecules self-assemble into elongated stacks that tend to collectively align along the same direction~\cite{zhou2017elasticity}, which we prescribe by rubbing the glass substrate with a diamond lapping film~\cite{suh2018nanoscratching}.
In this case, instead of becoming randomly-oriented, cells are retained end-to-end as they proliferate, forming single-cell-wide ``chains'' aligned along their long axis in the direction of LC alignment [Fig.~\ref{fig:fig_expt_chain}(b), first frame]. As proliferation progresses, the chains elongate, eventually buckling [second frame]---bending the surrounding LC in turn, as shown by the local reorientation of the LC director field $\boldsymbol{n}$ [Fig.~\ref{fig:fig_expt_chain}(c), Movie~S\ref{vid:dscg video growth}], which we visualize using crossed polarizers placed around the sample in the microscope's light path. Intriguingly, the chains form multiple, highly-localized, sharply-curved arcs as they buckle [arrows, second frame of Fig.~\ref{fig:fig_expt_chain}(b)--(c)]---in stark contrast to the classic Euler buckling instability of a compressed long, slender filament, which instead results in the formation of a single, long-wavelength arc. As the chains continue to elongate, these localized buckles continue to grow and new buckling events arise as well [Movie~S\ref{vid:dscg video growth}], eventually creating an intertwined three-dimensional network of long, serpentine, multicellular, single-cell-wide chains [Fig.~\ref{fig:fig_expt_chain}(d)].

To test the generality of this phenomenon, we perform the same experiment using nonmotile strains of two other species of bacteria: \textit{Vibrio cholerae} and \textit{Pseudomonas aeruginosa}. We again observe chain formation in both cases~[Fig.~\ref{fig:cholerae_pseudo_supp}(a)--(b)], indicating that this phenomenon arises across different cell types. Together, these results demonstrate that nonmotile bacterial colonies proliferating in nematic LCs generically form large, serpentine, single-cell-wide chains.


\begin{figure*}
\includegraphics[width=\textwidth]{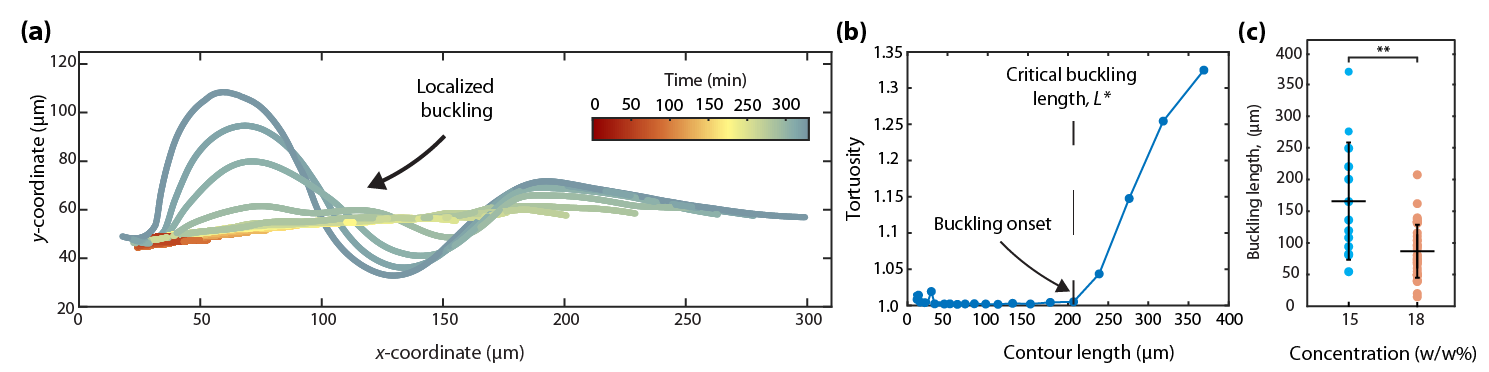}
\caption{\textbf{Characterization of chain buckling.} \textbf{(a)} Successive centerline configurations of a buckling chain in \SI{18}{w/w\%} DSCG, with colors indicating time. The chain initially grows straight before developing highly localized, sharply curved buckles. \textbf{(b)} Tortuosity (contour length divided by end-to-end distance) as a function of contour length for the chain in \textbf{a}; the sharp increase marks the onset of buckling at the critical length $L^*$. \textbf{(c)} Measured buckling lengths for chains in \SI{15}{w/w\%} (light blue, $n=13$) and \SI{18}{w/w\%} (light orange, $n=38$) DSCG. Horizontal lines indicate means; the critical buckling length decreases with increasing LC concentration (~**~indicates $p \leq 0.01$, two-sample $t$-test).
}
\label{fig:fig_chain_char} 
\end{figure*}

\textbf{Nematic elasticity causes chain formation.} Why do chains form? One possibility is that interactions with the surrounding LC alter how the individual cells grow, somehow causing them to proliferate in chains. However, bacterial growth rates directly measured using single-cell microscopy show no appreciable differences upon LC addition to the solution~[Fig.~\ref{fig:growth_curves}(A)--(B)]. Moreover, the osmotic pressure of the LC is $\approx\SI{2}{\kPa}$ [\emph{Supplementary Information}], far lower than the pressure at which bacterial physiology is typically altered,~$\SI{\sim100}{\kPa}$~\cite{rubinstein2012osmotic,buda2016dynamics,cayley_biophysical_2000,deng_direct_2011}---further arguing against this possibility.  

What other features of the LC drive chain formation? Close inspection of the LC director field $\boldsymbol{n}$ around a pair of cells during the nascent stage of chain formation provides a clue. The cells orient end-to-end along the imposed direction of LC alignment, with two opposing pairs of bright and dark regions flanking their outer poles apparent in the crossed polarizer micrograph [Fig.~\ref{fig:fig_expt_chain}(e)--(f)]. These features are known as ``boojums'', topological defects that form at the cell poles where the tangentially-anchored nematic director field must accommodate the curved spherocylindrical cell~{\cite{lavrentovich1998topological,pw98}}. 
The emergence of these point defects at the poles, rather than ring defects encircling the cell or complete disruption of the nematic order, indicates {that the LC is tangentially anchored to the cell surface with a finite anchoring strength}
~\cite{mushenheim2014dynamic,smalyukh_elasticity-mediated_2008}. 
That is, the elongated constituents of the LC tend to collectively align along both the direction imposed on the glass substrate as well as the surfaces of the cells. 
 Deviations from this preferred alignment cost energy; following the so-called one-constant approximation, all such deviations are penalized by a single Frank elastic constant, $K$. 
A minute rotation of the cell's long axis away from the overall imposed direction of the nematic director by {an angle} $\theta = 2^\circ$ requires $U_\mathrm{el} \approx 2 \pi (l_\mathrm{b}/\mathrm{log}\left(2l_\mathrm{b}/a\right))K\theta^2 \sim 500k_\mathrm{B}T$ in energy~\cite{crystal1987elastic,mushenheim_dynamic_2014,smalyukh_elasticity-mediated_2008}, where $l_\mathrm{b}$ and $a$ are the cell length and cross-sectional radius, respectively---far greater than the thermal energy $k_\mathrm{B}T$ [\emph{Supplementary Information}]. Therefore, the nematic elasticity of the LC forces cells to strongly align with it~\cite{smalyukh2008elasticity,mushenheim2014dynamic} and continue to proliferate as aligned chains. 

Two additional sets of experiments confirm this picture. First, we culture cells in nutrient-free solutions in which proliferation is arrested. When the solution does not contain LC, the cells are disconnected and randomly dispersed, with no orientational order [Fig.~\ref{fig:fig_lc_forces}(a)(i)]; by contrast, in the LC, the cells orient along the imposed direction of LC alignment [Fig.~\ref{fig:fig_lc_forces}(a)(ii)], consistent with our expectation and with previous observations of elongated ellipsoidal colloids in a nematic LC~\cite{tasinkevych2014dispersions}. Second, given that the nematic elasticity driving chain formation is generated by the LC, we expect that after a chain has formed, removing the LC from the solution around it should allow the chain to disintegrate via random thermal motion of the cells. This is precisely what we observe upon flushing a suspension of chains with LC-free liquid [Fig.~\ref{fig:fig_lc_forces}(b), Movie~S\ref{vid:dilution video}].  Taken altogether, these observations establish that nematic elasticity induced by the LC causes chains to form. 


\textbf{Chains buckle upon reaching a critical length.} Having established the origin of chain formation, we next ask: Why do chains buckle as their constituent cells continue to proliferate [Fig.~\ref{fig:fig_expt_chain}(b)--(c)]? And why does their buckling appear to be so different from the classic Euler buckling instability? To begin to address these questions, we first examine our experimental observations of chain buckling. Snapshots of the contour of a representative chain as it elongates are shown in Fig.~\ref{fig:fig_chain_char}(a); it {is initially straight}, but buckles, becoming more tortuous, after elongating sufficiently. Direct measurement of the chain's tortuosity, defined as its contour length $L$ normalized by its end-to-end distance, as it grows quantifies this observation: as shown in Fig.~\ref{fig:fig_chain_char}(b), the tortuosity increases sharply as the chain elongates beyond a length $\approx\SI{200}{\micro m}$. We {denote} this critical buckling length by $L^*$. Repeating this measurement for many other chains indicates that they consistently buckle at a critical length that decreases with increasing LC concentration (i.e., increasing nematic elasticity): $L^*=170\pm\SI{90}{\micro m}$ and $90\pm\SI{40}{\micro m}$ for experiments performed using $c=\SI{15}{w/w\%}$ and $\SI{18}{w/w\%}$, respectively [Fig.~\ref{fig:fig_chain_char}(c)]. In all cases, the buckling is far more localized than the long-wavelength arc formation characteristic of Euler buckling, as in Fig.~\ref{fig:fig_expt_chain}(b)--(c).

\begin{figure}
    \includegraphics[width=\linewidth]{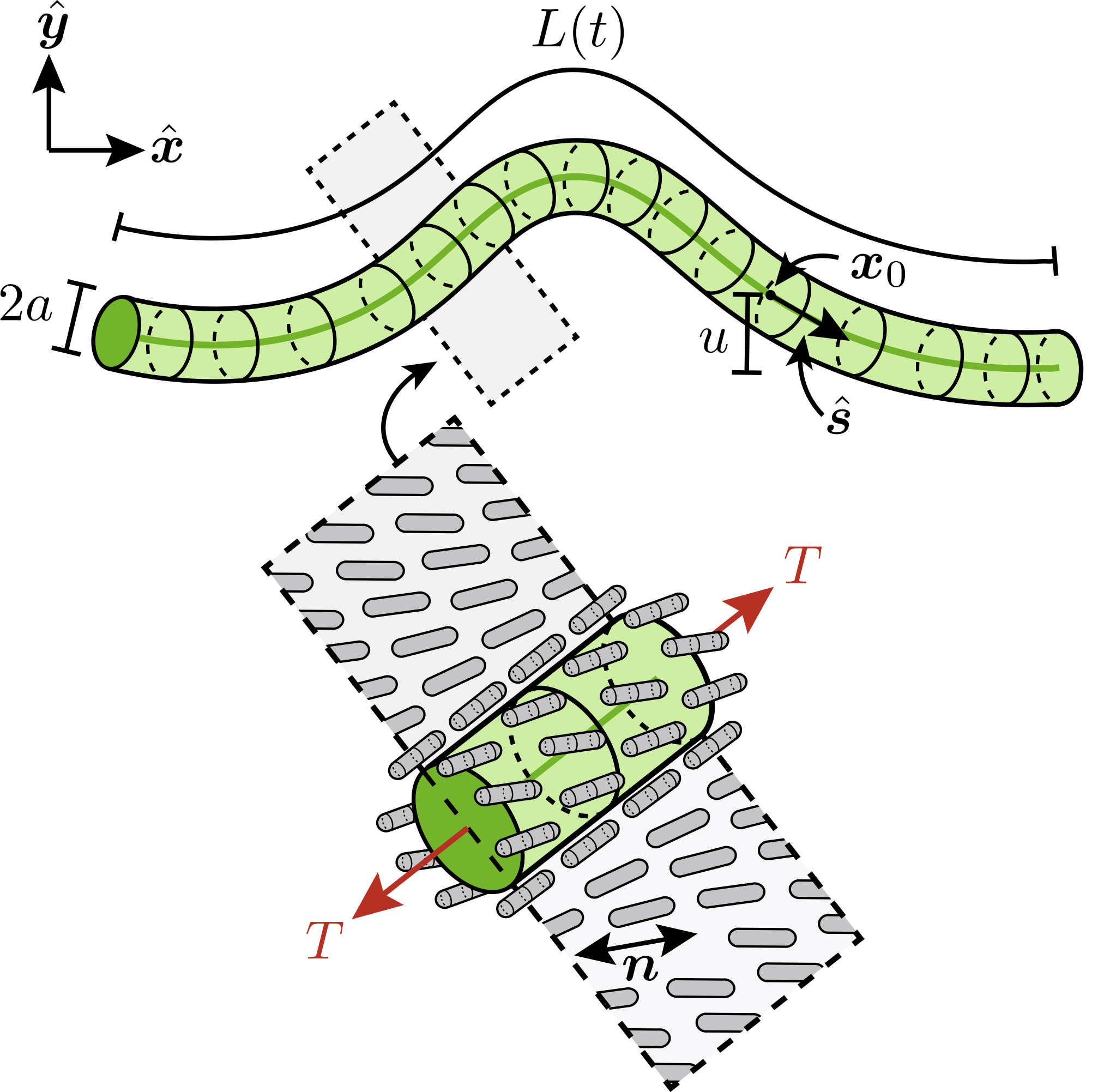}
    \caption{\textbf{Continuum model of a growing bacterial chain in a nematic LC.} The chain is treated as a slender filament of fixed radius $a$ and growing length $L(t)$. The transverse ($\hat{\b{y}}$) displacement $u(s,t)$ is parameterized by the arclength $s$. Growth generates internal compression $T(s,t)$ along the chain contour due to viscous drag from the surrounding LC, while chain deflections are resisted by the elastic restoring force from the deformed nematic director field (wavy lines, bottom).}
    \label{fig:schematic}
\end{figure}



\section{Mathematical Modeling}

\textbf{Force balance of a growing continuum chain.} Our experiments highlight the pivotal role of nematic elasticity in causing chains to form and influencing how they buckle. Indeed, while the bacterial chains are composed of discrete, unattached cells, the nematic elasticity of the surrounding LC effectively couples adjacent cells together---causing deformations of a given chain to be mechanically correlated over distances much larger than a single cell length. Hence, to shed light on the physics underlying chain growth and buckling, we turn to mathematical modeling. In particular, we describe each chain as a continuous, slender, elastic filament [Fig.~\ref{fig:schematic}] of fixed radius $a$ and a length that grows exponentially in time $t$, $L(t)=L(0)e^{\alpha t}$, where $\alpha$ is the single-cell growth rate. The chain centerline is parameterized by the arclength  $s\in[-L(t)/2,L(t)/2]$ as $\x_0(s,t)$, with centerline unit tangent vector $\shat = \partial_s\x_0$. The material velocity at each point, $\b{V}(s,t)$, can be decomposed as $\b{V}=\alpha s\shat+\b{v}(s,t)$, where the first term reflects tangential growth. 

The force per unit length acting on the chain at each station (point along the centerline) $s$ is then given by 
\begin{gather}\label{eq: force balance}
   \b{f}(s,t) = \f^{\text{viscous}}(s,t) +\f^{\text{LC}}(s,t)+\partial_s\left[T(s,t)\shat\right],
\end{gather}
where $\f^{\text{viscous}}$ is the viscous drag per unit length induced by the LC, $\f^{\text{LC}}$ is the elastic restoring force per unit length from the deformed nematic, and $T$ is the internal tension (or compression if $T<0$) arising from bacterial growth. Because viscous damping is much larger than inertial effects in our experiments, $\b{f} \equiv\b{0}$ at all times \cite{Graham18}; since the bacterial chain grows very slowly compared to the other time scales governing the LC flow, the LC is assumed to be in quasi-static equilibrium at all times. Moreover, because the chain is slender, we can use the resistive force approximation, which assumes the drag at each point depends only on the local chain velocity and orientation, to calculate $\f^{\text{viscous}} =-\left[\zeta_{\parallel}\shat\shat+\zeta_{\perp}\left(\I-\shat\shat\right)\right]\cdot \bm{V}$; the drag coefficients $\zeta_{\parallel}$ and $\zeta_{\perp}$ penalize motion in the chain tangent and normal directions, respectively. Eq.~\eqref{eq: force balance} can then be decomposed into tangential and normal components as
\begin{subequations}
\begin{gather}
    -\zeta_{\parallel}(v_{\parallel}+\alpha s) +\partial_s T=0,\label{eq: force_balance_tangential}\\
    -\zeta_{\perp}\v_{\perp} + (\I-\shat\shat) \cdot \left(\b{f}^{\text{LC}}+T\partial_s\shat\right)=\bm{0}, \label{eq: force_balance_normal}
\end{gather}
\end{subequations}
where the material velocity $\v=v_{\parallel}\shat+\v_{\perp}$ with $\shat\cdot\v_{\perp}=0$, and $\shat \cdot \b{f}^{\text{LC}}=0$ [detailed in \emph{Supplementary Information}]. Assuming small chain displacements for simplicity, we approximate the deformed position as $\x_0(s,t)\approx s\xhat+u(s,t)\yhat$, 
where $\xhat$, $\yhat$, and $\zhat$ denote the unit Cartesian vectors and $u$ is the vertical displacement. We then have that $\v_\perp \approx \partial_tu \yhat$ and $\shat \approx \xhat+\partial_su \yhat$, 
giving 
\begin{equation}
\zeta_{\perp}\partial_tu =T(s,t)\partial_{ss}u+\yhat \cdot \b{f}^{\text{LC}}(s,t).\label{eq: filament_dynamics}
\end{equation}

To close this system of equations, we compute the tension $T$ and LC restoring force $\bm{f}^{\text{LC}}$ for the two principal cases that arise in the experiments: when the chain is (i) free at both ends (``free-free'') or (ii) fixed at one end, e.g., due to pinning on the glass substrate and free at the other (``fixed-free''). Details are provided in the \emph{Supplementary Information}. 

In brief, integrating Eq.~\eqref{eq: force_balance_tangential} yields the tension $T(s,t)$, which depends on $\alpha$, $\zeta_{\parallel}$, $L(t)$, and the endpoint tension $T_L$. In both cases (i) and (ii), as growth progresses, the tension eventually becomes negative (i.e., compressive)---introducing the possibility of buckling~\cite{Vella2019}. However, unlike in classical Euler buckling, the load is distributed \emph{along} the length of the chain due to the viscous drag, just as in the buckling of sedimenting filaments ~\cite{lmss13} {and filaments placed in compressional flows \cite{ms15,dlns19,cllcfdsl20}}. To compute the LC elastic restoring force, we assume that the LC is deep in the nematic state with director field $\n(\x,t)$ and $|\n|=1$ \cite{dp93}. With $\n\to \xhat$ as $|\x|\to\infty$, we assume for simplicity that $\n(\x)\approx \xhat+\b{p}(\x)$ with $\xhat\cdot\b{p}=\bm{0}$, for a small perturbation in the molecular orientation $\b{p}=\Oh(\partial_su)$. Using the one-constant approximation for the Frank elastic energy, $\nabla^2\b{p}=\bm{0}$ \cite{dp93}, enabling us to compute the LC traction (surface force per unit area) acting on the chain. We thereby compute $\bm{f}^{\text{LC}}(s,t)$, which depends on $K$, $a$, $L(t)$, and the anchoring strength $W$ quantifying the energetic penalty for LC molecules to deviate from tangential alignment with the chain surface.

\textbf{Parameters characterizing bacterial chain growth in a liquid crystal.} To distill the relevant physics from our theoretical model, we first reduce the number of parameters in Eqs.~\eqref{eq: force balance}--\eqref{eq: filament_dynamics} by scaling lengths and time by the constant chain radius $a$ and inverse growth rate $\alpha^{-1}$, respectively. Using these natural scales, we obtain four key dimensionless parameters that govern chain growth:
\begin{itemize}
\item The chain aspect ratio, $\Lambda \equiv L/(2a)$, which increases to $\gg1$ as growth progresses.
\item The drag anisotropy ratio, $\chi\equiv\zeta_{\perp}/\zeta_{\parallel}$. Approximating the chain as a filament moving in an isotropic fluid a distance $h$ from a wall, with $h\ll L$ \cite{wigg19}, $\zeta_\parallel\approx[2\pi/\log(2h/a)] \mu_\parallel$ and $\chi\equiv\zeta_\perp/\zeta_\parallel\approx 2\mu_\perp/\mu_\parallel$ for the LC perpendicular and parallel viscosities $\mu_\perp$ and $\mu_\parallel$, respectively. Typical viscosity ratios for DSCG range from $\mu_\perp/\mu_\parallel\approx2$ to $6$  \cite{duchesne2015bacterial,Gomez2016}, so we fix $\chi\approx5$.
\item The dimensionless anchoring strength, $w\equiv aW/K$. We estimate that our experiments probe $w\sim10^{-2}$--$10^{0}$~\cite{mushenheim_dynamic_2014,zhou2017dynamic,chi2020surface} (\emph{Methods}), consistent with the weak tangential anchoring indicated by the micrograph in Fig.~\ref{fig:fig_expt_chain}{(f)}.
\item The Ericksen number, $\Er\equiv a^2\zeta_\parallel\alpha /(2\pi K)$, which compares viscous forces introduced by chain growth to elastic forces arising from LC deformation. We estimate (\emph{Methods}) that our experiments probe $\Er\approx 3.0\times 10^{-7}$ ($c=\SI{15}{w/w\%}$) and $7.7\times 10^{-7}$  ($c=\SI{18}{w/w\%}$){. $\Er\ll1$}, which highlights the dominant role of nematic LC elasticity in controlling chain growth. 
\end{itemize}
As detailed in the \emph{Supplementary Information}, we neglect the endpoint tension $T_L$ because it is orders of magnitude smaller than the viscous stress $a^2 \zeta_{\parallel} \alpha$, thereby simplifying our system of equations further.



\textbf{Competition between LC elasticity and growth-induced stresses along the chain contour drives localized buckling.} Having nondimensionalized the governing equations, we predict if and when a growing chain will buckle by performing a linear stability analysis of the straight chain state [detailed in the \emph{Supplementary Information}]. That is, we ask: at what chain length (or equivalently, aspect ratio $\Lambda$) does a small perturbation to the straight configuration begin to grow rather than decay? To address this question using our model, we perturb the chain centerline with a small displacement $u=a e^{\sigma\tau}U(\xi)$ [Fig.~\ref{fig:schematic}], where $\xi\equiv2s/L(t)\in[-1,1]$ and $\tau \equiv  \alpha t$; here, $\sigma$ is the growth rate of the perturbation. We then expand the non-dimensional displacement $U(\xi,\tau)$ as a Fourier series with time-dependent coefficients and substitute into the dimensionless form of Eq.~\eqref{eq: filament_dynamics}, with $\Lambda(\tau)$ assumed to be frozen at its instantaneous value $\approx \Lambda$. This procedure yields an eigenvalue problem for the growth rate $\sigma_k$ of each Fourier mode $\hat{U}_k$ as a function of $\Lambda$:
\begin{equation}\label{eq:eigen}
    \chi \sigma_k \hat{U}_k = \left(\frac{\chi}{2}+\frac{q_k^2}{3}-\frac{1}{4}-\frac{q_k^3}{\Er\cdot\Lambda^3 Z_k}\right)\hat{U}_k+\sum_{m\neq k} J_{km} \hat{U}_m,
\end{equation}
for the free-free case, where $Z_k$ quantifies how strongly the LC resists chain deformations at mode $k$ and $J_{km}$ are coupling coefficients between modes, both derived in the \emph{Supplementary Information}. The fixed-free case yields an analogous eigenvalue problem with modified mode functions and coupling coefficients. Chain buckling then occurs when any of the modes becomes unstable ($\sigma_k>0$), which defines the critical buckling aspect ratio $\Lambda^*$ corresponding to the critical length $L^*$.

\begin{figure}[ht!]
    \centering
    \includegraphics[width=\linewidth]{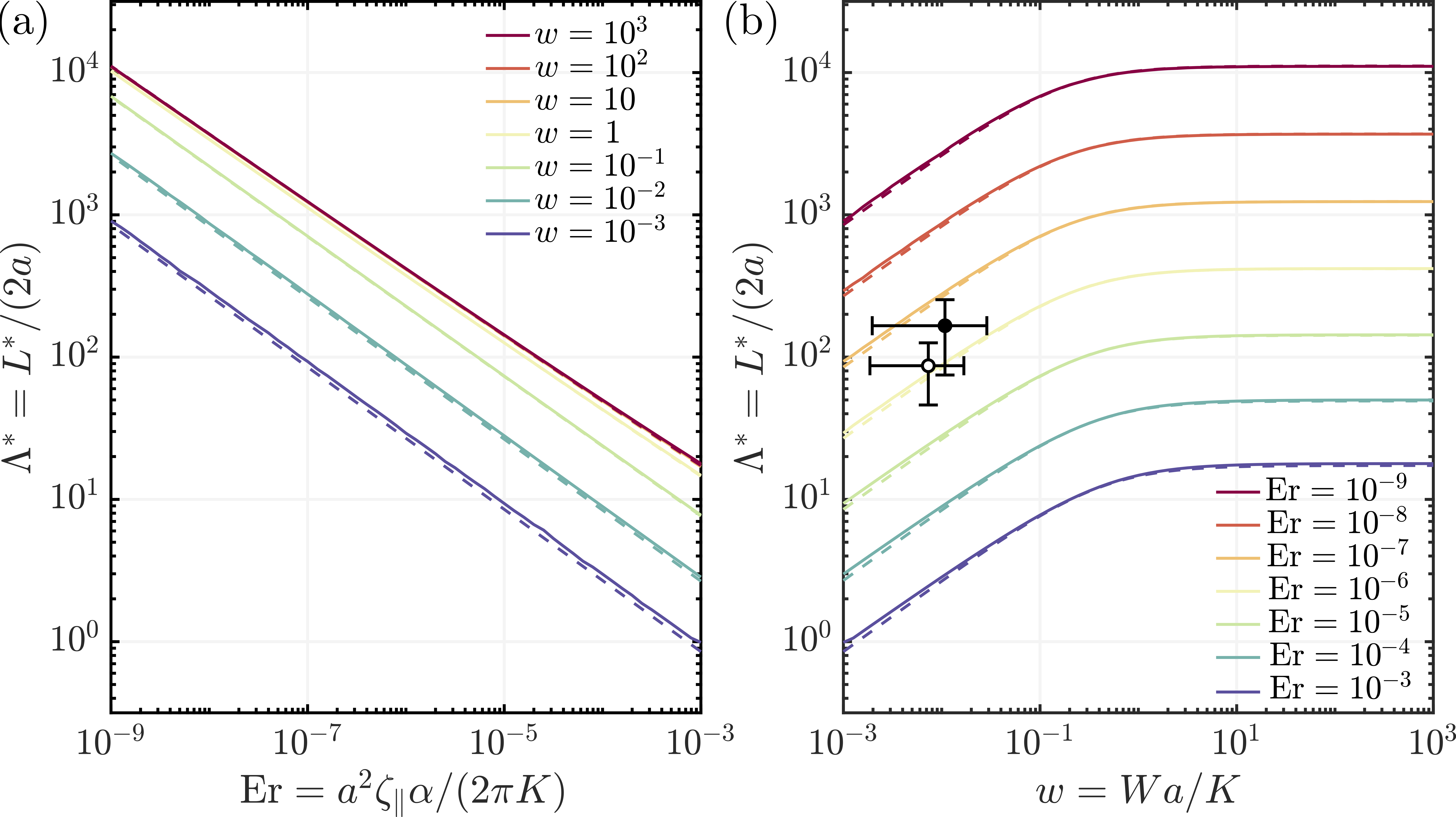}
    \caption{\textbf{Theoretical predictions for the critical buckling length.} \textbf{(a)} Dimensionless critical buckling length $\Lambda^*\equiv L^*/(2a)$ as a function of the Ericksen number $\Er\equiv a^2\zeta_\parallel\alpha /(2\pi K)$ for different values of the dimensionless anchoring strength $w\equiv aW/K$. Solid curves show numerical solutions of the linear stability analysis [Eq.~\eqref{eq:eigen}]; dashed curves show the analytical approximation [Eq.~\eqref{eq:fittedmixed}]. The scaling $\Lambda^*\sim\Er^{-1/2}$ reflects the balance between growth-induced viscous compression and LC elastic restoring forces. \textbf{(b)}  $\Lambda^*$ as a function of $w$ for different values of $\Er$. For weak anchoring ($w\ll1$), the director can slip relative to the bacterial cell surface, reducing the effective elastic coupling and allowing buckling at shorter chain lengths; for strong anchoring ($w\gg1$), the coupling saturates and $\Lambda^*$ becomes independent of $w$. Both panels show predictions for the free-free boundary conditions and $\chi\approx5$ [\emph{Methods}].
    Symbols show experimentally measured buckling lengths at 15 w/w\% (filled circle) and 18 w/w\% (open circle) DSCG, plotted at their corresponding $\Er$ values; comparison with theory yields inferred anchoring strengths $w\approx0.01$, corresponding to $W\approx 2.6\times 10^{-7}\,\si{\J/\meter^2}$.
    }
    \label{fig:Lstar}
\end{figure}

Our numerical solutions of $\Lambda^*$ from Eq.~\eqref{eq:eigen} are shown in Fig.~\ref{fig:Lstar} (solid curves) for the exemplary free-free case. The results for the fixed-free case are similar in magnitude, but shifted slightly lower by a factor of $\approx1.5$, because a fixed end cannot relieve growth-induced compression by moving outward, unlike a free end---which is reflected in differences in $T(s,t)$ between the two cases [\emph{Supplementary Information}]. Inspired by the solution when the Fourier modes are decoupled ($J_{km}=0$), we also approximate our results as 
 \begin{gather}\label{eq:fittedmixed}
   \Lambda^*  \approx \sqrt{\displaystyle\frac{a_1\Er^{-1}}{\log (a_2 \Er^{-1})+4/w}},
 \end{gather}
which provides a useful analytical expression for $\Lambda^*(\Er,w)$, shown by the dashed curves; here, $\{a_1,a_2\}$ are numerical constants equal to $\{2.91,16.2\}$ and $\{1.26,0.664\}$ for the free-free and fixed-free cases, respectively. 

To test our theoretical predictions, we compare the measured critical buckling lengths [Fig.~\ref{fig:fig_chain_char}(c)] with those predicted by our theory. As shown by the symbols in Fig.~\ref{fig:Lstar}(b),  which indicate our measured values at their corresponding Ericksen numbers ($\Er\approx 3.0\times 10^{-7}$ and $\approx 7.7\times 10^{-7}$, respectively), we find good agreement with the free-free theoretical predictions with dimensionless anchoring strengths $w\approx 1.1\times 10^{-2}$ and $0.73\times 10^{-2}$, respectively; comparing to the fixed-free prediction instead yields slightly larger values, $w\approx 3.0\times 10^{-2}$ and $2.0\times 10^{-2}$, respectively. These values correspond to dimensional anchoring strengths $W\approx 2.6-2.8\times 10^{-7}\,\si{\J/\meter^2}$ and $7.2-7.6\times 10^{-7}\,\si{\J/\meter^2}$, respectively; because our experiments likely include chains with a mixture of boundary conditions, the true anchoring strength likely lies within these ranges [\emph{Supplementary Information}]. Nevertheless, these values are in excellent agreement with previous studies of bacteria and similar inclusions in nematic LCs, which found $W\approx 10^{-7}-10^{-5}\,\si{\J/\meter^2}$~\cite{mushenheim_dynamic_2014,zhou2017dynamic,chi2020surface}. 

The good quantitative agreement between theory and experiment validates our central physical picture: that the competition between LC nematic elasticity and growth-induced viscous stresses along the chain contour drives buckling. Fig.~\ref{fig:Lstar} reveals how these two effects control the critical buckling length. Consider first the dependence on the Ericksen number, $\Er\equiv a^2\zeta_\parallel\alpha /(2\pi K)$. The compressive viscous drag force $\sim \zeta_\parallel\alpha L^2$ increases as the chain grows, eventually exceeding the critical LC elastic restoring force $\sim K$ at $L=L^*$. Balancing these forces at buckling yields $\Lambda^*\sim\Er^{-1/2}$, as shown by the curves in Fig.~\ref{fig:Lstar}(a); larger $\Er$ (faster growth or weaker LC elasticity) leads to buckling at shorter chain lengths. At the small Ericksen numbers relevant to our experiments ($\Er\sim10^{-7}$), LC elasticity dominates, permitting chains to grow to hundreds of cells before the accumulated compression triggers a buckling instability. Next, consider the dependence on anchoring strength. For weak anchoring ($w\ll1$) such as in our experiments ($w\sim10^{-2}$), the director can slip relative to the bacterial cell surface, reducing the effective elastic coupling between chain deflections and the surrounding LC. This weakened coupling lowers the restoring force resisting buckling, which thus occurs at a relatively low accumulated compressive stress and thus for shorter chains---hence, in this regime, $\Lambda^*$ grows with $w$, as shown in Fig.~\ref{fig:Lstar}(b). By contrast, for strong anchoring ($w\gg1$), the LC director couples to the bacterial surface tightly and the elastic coupling saturates; further increases in $w$ provide no additional coupling, so $\Lambda^*$ becomes independent of $w$, as also shown in [Fig.~\ref{fig:Lstar}(b)]. 

The theoretical analysis also explains why chain buckling differs so strikingly from classical Euler buckling. In Euler buckling, an end-applied compressive load destabilizes a single, long-wavelength mode spanning the entire filament. For a bacterial chain in an LC, by contrast, compression is not applied at the endpoints but is generated by viscous drag continuously along the chain as it grows. Simultaneously, the surrounding nematic acts as a distributed elastic foundation, penalizing deflections at all wavelengths. Together, these effects promote instability at shorter wavelengths, producing the highly localized, sharply curved buckles observed in our experiments rather than a single gentle arc. Some experimental realizations, like that shown in Fig.~\ref{fig:fig_chain_char}, exhibit larger-wavelength deformations reminiscent of the wrinkling-to-folding transition in elastic media under large compressive loads~\cite{hc10}; exploring this connection further using a fully nonlinear theory will be an interesting direction for future work.

\section{Discussion}
By combining experiments and continuum mechanical theory, we have shown that liquid crystalline environments can shape bacterial colony morphology through purely physical interactions. Nonmotile bacteria proliferating in a nematic LC form long, serpentine, single-cell-wide chains aligned along the director---a morphology that arises because LC elasticity keeps proliferating cells oriented end-to-end. As these chains elongate, growth-induced viscous stresses accumulate along their contours, eventually overcoming the elastic penalty imposed by the surrounding nematic and triggering highly-localized buckling that starkly contrasts to the long-wavelength bending typical of classical Euler buckling---conceptually similar to how the elastic cytoskeleton suppresses bending of microtubules inside cells~\cite{brangwynne2006microtubules}. In addition to {explaining} all our experimental observations, our theoretical analysis provides quantitative principles that describe the dependence of chain buckling on the Ericksen number and LC--cell surface anchoring strength, motivating future studies that explore variations in these parameters further. Indeed, while our theory quantitatively captures the decrease in critical buckling length with increasing LC concentration, the spread in measured $L^*$ values suggests additional variability---likely from cell-to-cell differences in anchoring strength, and possibly from imperfection sensitivity or subcritical effects that could be disentangled through analysis of a fully nonlinear model.

{Intriguingly,} bacterial colony morphologies reminiscent of the chains studied here have been observed in various biological contexts. However, their origin is typically attributed to specific adhesive interactions holding cells together~\cite{moor_high-avidity_2017,mishra_mechanopathology_2023,azimi2021specific,puri_evidence_2023}. Here, we reveal a fundamentally distinct mechanism by which proliferating bacteria can form such long, single-cell-wide chains: through mechanical interactions with a surrounding liquid crystal. This finding, which complements our previous study of ``cables'' formed by cells proliferating in polymer solutions~\cite{gonzalez2025morphogenesis}, highlights how purely physical interactions with surrounding complex fluids---such as host mucus~\cite{figueroa2019mechanical,viney1993liquid,davies1998water}, extracellular polymeric substances in biofilms~\cite{repula2022biotropic}, and dense suspensions of filamentous phages~\cite{secor2015biofilm,secor2015filamentous,tarafder_phage_2020}---can profoundly shape growing bacterial colonies.

What are the biological implications of chain formation? Our findings suggest that LC-mediated chain formation could influence how nonmotile bacteria explore space, access nutrients, and interact with external stressors. Long chains have large surface-area-to-volume ratios, potentially increasing susceptibility to antibiotics or phages~\cite{slk90,hh99,psvpssp99,cj06,tarafder_phage_2020,secor2015biofilm,secor2015filamentous}; alternatively, the extended chain morphology could enable colonies to reach new environmental niches~\cite{mishra_mechanopathology_2023,losick2020bacillus} or nutrient sources~\cite{Young06,Young07}. The buckling instability itself may serve as a mechanical signal that triggers phenotypic changes in the cells themselves; indeed, our theory suggests that the compressive force in chains reaches $\SI{\sim1}{pN}$ at the onset of buckling---comparable to the forces that drive bacterial mechanosensitive responses~\cite{chawla2020skeptic}, such as biofilm formation. Investigating these possibilities, both in quiescent conditions and under flow~\cite{gmbpbsg20,ebbm89}, will be an important direction for future work.

Our work also opens new directions in soft matter physics. While colloidal assembly in LCs has been extensively studied~\cite{tasinkevych2014dispersions,mondiot2014colloidal,sudha2021colloidal,zimmermann2015self}, our results highlight that qualitatively new behaviors emerge when the constituent particles can proliferate~\cite{hallatschek2023proliferating}. The interplay between growth-generated stresses, LC elasticity, and surface anchoring creates a rich design space for controlling colony morphology, and potentially for engineering living materials with prescribed architectures. Extending our framework to other anisotropic environments, such as cholesteric LCs or nematic gels with tunable elasticity, could reveal additional morphogenetic programs accessible through physical interactions alone~\cite{trivedi2016nonsingular}.

\begin{acknowledgments}
We thank Z.~Gitai and J.~Shaevitz for useful discussions, and the laboratories of B.~Austin, B.~Bassler, and Z.~Gitai for providing strains of \emph{E.~coli}, \emph{V.~cholerae}, and \emph{P.~aeruginosa}, respectively. \\

\textbf{Funding}: S.E.S.\ acknowledges support from NSF grant DMS-2527011. N.S.W.\ acknowledges support from NSF Center for the Physics of Biological Function grant PHY-1734030. This research was also supported by the National Institute of General Medical Sciences (NIGMS) of the National Institutes of Health (NIH) under Award Number R01 GM082938. The content is solely the responsibility of the authors and does not necessarily represent the official views of the National Institutes of Health. S.S.D. acknowledges support from NSF grants CBET-1941716, DMR-2011750, and EF-2124863 as well as the Camille Dreyfus Teacher-Scholar and Pew Biomedical Scholars Programs.\\

\textbf{Author contributions}: S.G.L.C.\ and S.S.D.\ designed the overall research project; S.G.L.C.\ performed all experiments and experimental analyses; T.G.J.C.\ and S.E.S.\ developed the theoretical model and performed all calculations; S.G.L.C., T.G.J.C., S.E.S., N.S.W., and S.S.D.\ analyzed data and wrote
the paper.\\

\textbf{Competing interests}: The authors declare that they have no competing interests.
\end{acknowledgments}

\section{Methods}

\subsection{Preparation of cells}
\subsubsection{\textit{Escherichia coli}}
We incubated an overnight culture of \textit{E.~coli} strain W3110 in \SI{2}{w/w\%} Lennox Lysogeny Broth (LB) at \SI{30}{\degreeCelsius}. Next, we inoculated \SI{20}{\uL} from the overnight culture in \SI{2}{\mL} of LB for 3 hours  such that, at the time of imaging, bacteria would reach mid-exponential phase.  The \textit{E.~coli} constitutively express green fluorescent protein (GFP) throughout their cytoplasm and have a deletion of the flagellar regulatory gene \textit{flhDC} that renders them nonmotile. 

\subsubsection{\textit{Pseudomonas aeruginosa}}
We incubated an overnight culture of \textit{P. aeruginosa} strain PA01 in LB, supplemented with carbenicillin (\SI{200}{\ug/\mL}) to preserve the strain’s fluorescence, at \SI{37}{\degreeCelsius}. Next, we inoculated \SI{20}{\uL} from the overnight culture in \SI{2}{\mL} of LB for 3 hours. We used a strain with a double \textit{fliC} and \textit{pilA} deletion that renders the cells nonmotile, and verified that these cells are nonmotile using direct visualization. 

\subsubsection{\textit{Vibrio cholerae}}

We incubated an overnight culture of \textit{V. cholerae} 01 biovar El Tor strain C6706 in LB at \SI{37}{\degreeCelsius}. Next, we inoculated \SI{20}{\uL} from the overnight culture in \SI{2}{\mL} of LB for 3 hours at \SI{37}{\degreeCelsius}. We used a strain with several gene deletions: It has a deletion of \textit{pomA}, which renders the cells nonmotile, and we verified that these cells are nonmotile using direct visualization. It also has deletions of \textit{rbmA}, \textit{bap1}, \textit{rbmC}, and \textit{vpsL}, which renders the cells as nonbiofilm formers. 

\subsection{Preparation of liquid crystal solution}
We purchased liquid crystal disodium chromoglycate (DSCG) from Thermo Fisher Scientific. DSCG stock solutions were prepared in LB and put in a spinning rotor until full dissolution was observed. The DSCG solution was then filtered using a \SI{5}{\um} membrane from GE Healthcare life sciences. To obtain liquid crystal solutions at the desired concentrations we diluted the stock solutions with the same solvent. 

\subsection{Imaging of bacteria in liquid crystal solution} 

\subsubsection{\textit{E.~coli} in liquid crystal}

To image fluorescent \textit{E.~coli} we used custom built rectangular PDMS devices  with dimensions \SI{2}{\mm} in width,  \SI{22}{\mm} in length, and  \SI{25}{\um} in height. The device was filled with the liquid crystal solution containing a well mixed bacterial population at a concentration~$c_\mathrm{cell}~\approx 10^5$--$\SI{e6}{cells/\mL}$. Next, we sealed inlet and outlet ports of the PDMS channel with paraffin oil to minimize evaporation. Finally, we imaged the dish with a $20\times$ magnification air objective [Fig. \ref{fig:fig_expt_chain}(a)--(c)] or $60\times$ magnification oil objective with added $1.5\times$ zoom [Fig. \ref{fig:fig_expt_chain}(f)] in a Nikon AXR inverted laser-scanning confocal microscope with the stage maintained at \SI
{30\pm 1}{\degreeCelsius}. We recorded images throughout the sample every 1--10~min over $\sim 6$--$14$ hours. To obtain the images in Fig. \ref{fig:fig_expt_chain}(c),(f), we inserted a linear polarizer on the lightpath towards the transmittance detector.

To investigate the reversibility of chain formation [Fig.~\ref{fig:fig_lc_forces}B], we used a transparent-walled glass-bottom petri dish sealed with an overlying PDMS slab. We additionally pierced the PDMS slab with three 20-gauge needles, sealed using optical glue, and connected to a Harvard Apparatus 11 Elite syringe pump to provide fluid in/outflow. In particular, one needle acts as an inlet for injection of the test solution containing~$\approx \SI{4e5}{cells/\mL}$ suspended in \SI{15}{w/w\%} DSCG solution; the second acts as an inlet for injection of DSCG-free LB media; and the third acts as the outlet. We then imaged the sealed chamber from below using a Nikon A1R+ inverted laser scanning confocal microscope with the stage maintained at \SI
{30\pm 1}{\degreeCelsius} as before, first under quiescent no-flow conditions with the chamber filled with the test solution containing cells and LC. Once chains have formed, we then pump LC-free fluid through at a flow rate of \SI{2.5}{\uL/\min} to remove LC-containing solution.

To determine the onset of chain buckling [Fig.~\ref{fig:fig_chain_char}], we (i) smooth the confocal micrographs of growing chains using a 2D Gaussian filter, (ii) binarize the images using Otsu’s method of thresholding~\cite{sezgin2004survey}, (iii) skeletonize the individual chains using the medial axis transform and thereby determine their backbones, and (iv) directly measure the tortuosity of each identified chain backbone.

\subsubsection{\textit{P. aeruginosa} and \textit{V. cholerae} in liquid crystal}

We filled individual wells of a glass-bottom 96-well plate with 100~$\mu$L of the solution and $0.1~\mu\mathrm{L}$ of the inoculum of cells such that their initial concentration is $\approx\SI{e6}{cells/\mL}$. We stored the plate in a static $\SI{30}{\degreeCelsius}$ incubator and imaged the plate at different time points: For \textit{P. aeruginosa}, we imaged at 120 and 840 min after the start of incubation, and for \textit{V. cholerae}, we imaged at 180 and 360 min after incubation as shown in Fig.~\ref{fig:cholerae_pseudo_supp}A-B. For all images taken, we used a $20\times$ air objective mounted in a Nikon A1R+ inverted laser scanning confocal microscope with the stage maintained at room temperature.\\

\renewcommand{\thefigure}{S\arabic{figure}}
\renewcommand{\thetable}{S\arabic{table}}
\renewcommand{\thesection}{S\arabic{section}}

\newpage
\onecolumngrid
\setcounter{figure}{0}
\setcounter{table}{0}
\setcounter{equation}{0}
\section*{Supplementary Information}

\subsection*{Osmotic pressure of nematic liquid crystal solutions}

To estimate the osmotic pressure of nematic liquid crystals (LCs) we use the expression~\cite{van2000depletion} 
\begin{equation}
    \Pi_\mathrm{osm} = 3n_\mathrm{p} k_\mathrm{B}T,
\end{equation}
where $n_\mathrm{p}$ is the number density of nematic liquid crystal molecules, $k_\mathrm{B}$ is Boltzmann's constant, and $T$ is the absolute temperature. We estimate the number density as $n_\mathrm{p} \approx 4\phi/(L^2 D \pi)$~\cite{van2000depletion,lekkerkerker_colloids_2011}, where $\phi$ is the fractional volume occupied by the LC particles, $D$ is the average diameter of the LC particles, and $l$ is the average length of the LC particles. Inputting the values of Table~\ref{tab: table_expts_th} renders $\Pi_\mathrm{osm} =\SI{2000}{\Pa}$.

\subsection*{Estimation of chain distance from a surface}
We observe that buckling occurs in chains that have at least one endpoint transiently pinned,  suggesting that bacterial chains come into contact with the surfaces of the imaging device (the chain's endpoints may come into contact with the glass bottom surface~\cite{vissers2018bacteria} or with the top PDMS surface~\cite{stepulane2022multifunctional}). To incorporate these surface effects into the theoretical calculations, we assume that the filament is growing a distance, $h$, from a surface~(Fig.~\ref{fig:h schematic}). To estimate $h$ experimentally, we track the $z$-coordinate of a chain's endpoints throughout the experiment. Next, we calculate the standard deviation of the $z$-coordinate time series. The standard deviation yields an estimate of the chain's fluctuations near the surface of the imaging device, providing an estimate of a chain's average distance from the surface. We repeated this procedure for multiple buckling chains. The averages  of these measurements are shown in Table~\ref{tab: table_expts_th} for different DSCG concentrations.

\section*{Mathematical modeling}

\subsection{Slender filament growth and kinematics}\label{sec:kinematics}

The bacterial chain will be treated as a continuum filament of length $L(t)$, which starts at $t=0$ in a straight configuration with centerline $\X_0(S)$ parameterized by the arclength $S\in[-L(0)/2,L(0)/2]$ (the reference frame). At time $t$, the centerline position is written as $\x_0(s(S,t),t)$, where $s\in[-L(t)/2,L(t)/2]$ is the current arclength parameter and $|\partial_s\x_0|=1$.

Assuming a uniform growth law for each unit of bacterial length, we have $\partial s/\partial S = \alpha$ and $\dot{s}\equiv\partial_t s(S,t)=\alpha s(S,t)$. Hence, with $s(S,0)=S$, we have naturally that $s(S,t)=S \exp(\alpha t)$. 

The material velocity, denoted by $\V(S,t)$, of the filament centerline is given by the time rate of change of the current position for fixed material coordinate $S$:
\begin{gather}
    \V(S,t) = \frac{\dd}{\dd t}\x_0(s(S,t),t)\Big|_{S} = \partial_t\x_0\Big|_{s}+ \dot{s}\,\partial_s\x_0 \eqqcolon\v(s(S,t),t)+\alpha s(S,t) \shat(s(S,t),t),
\end{gather}
where  $\shat\equiv\partial_s\x_0(s(S,t),t)$ is the unit tangent vector along the centerline. We recognize the final two terms as the rigid body motion component and the growth component, respectively. We further decompose the rigid body motion into its tangential and normal parts. At time $t$, 
\begin{gather}
    \v(s,t) = v_{\parallel}(s,t)\shat(s,t) + \v_\perp(s,t),
\end{gather}
where $\shat \cdot \v_\perp=0$. Continuity demands that $|\partial_s\x_0|=1$ at all times, which implies that
\begin{gather}
0=\frac{1}{2}\frac{\dd}{\dd t}|\partial_s\x_0|^2=\partial_s\x_0 \cdot \partial_{st}\x_0 = \shat \cdot \partial_s (\partial_t\x_0) = \shat \cdot \partial_s \v.
\end{gather}
As expected, the rigid body component of the filament velocity must be constant in the tangential direction, $v_{\parallel}(s,t)=v_{\parallel}(t)$, or else the filament would rupture.

\subsection{Filament geometry}
For a small planar displacement of the filament, $u$, we write the centerline position as
\begin{gather}
\b{x}_0(s,t)=s\xhat+u(s,t)\yhat+\Oh(L u_s^2),
\end{gather}
where $\{\xhat,\yhat,\zhat\}$ denote the unit Cartesian vectors, with $\xhat$ pointing along the length of the undeformed filament.
To study the initial instability we will assume that $|u_s|\ll 1$. The centerline unit tangent vector, $\b{\hat{s}}=\partial_s \x_0$, is accompanied by normal and binormal unit vectors along the centerline, $\{\nhat, \zhat\}$, where
\begin{subequations}
\begin{gather}
\shat=\xhat+u_s \yhat+\Oh(u_s^2),\\
\nhat =-u_s\xhat+ \yhat+\Oh(u_s^2),
\end{gather}
\end{subequations}
forming an orthonormal triad.

Each cross-section of the filament is assumed to be circular with diameter $2a$. The filament surface, denoted by $\partial D$ (neglecting the filament ends), is then parameterized by $(s,\eta)\in[-L(t)/2,L(t)/2]\times [0,2\pi)$, as
\begin{gather}
\x(s,\eta)=\x_0(s)+ a  \hat{\b{e}}_\perp(s,\eta),
\end{gather}
where $\hat{\b{e}}_\perp(s,\eta) = \cos\eta\, \nhat(s)+\sin\eta\zhat$. At each point on the surface, $\x(s,\eta)\in \partial D$, we have another orthonormal triad $\{\xihat,\zehat,\bnu\}$, composed of two unit tangent vectors on the surface,  $\xihat=\bm{x}_s/\lvert\bm{x}_s\rvert$ and $\zehat=\bm{x}_\eta/\lvert\bm{x}_\eta\rvert$, and the unit normal vector, which points into the LC, $\bnu=\zehat\times\xihat/\lvert\zehat\times\xihat\rvert$. With the centerline displacement assumed to be small, these are given by
\begin{subequations}
\begin{align}
     \bm{x}(s,\eta)&=s\xhat+a\rhat+\Oh(Lu_s),\label{eq:xpos_asym}\\
     \zehat(s,\eta)&=-\sin\eta \nhat(s) +\cos\eta \zhat+\Oh(u_s^2),\\
     \xihat(s,\eta)&=\shat(s)+\Oh(u_s^2),\\
    \bnu(s,\eta)&=\cos\eta \nhat(s) +\sin\eta \zhat+\Oh(u_s^2),
\end{align}
\end{subequations}
where  $\rhat(\eta)=\cos\eta \yhat+\sin\eta\zhat$ and $\bbeta(\eta)=-\sin\eta \yhat+\cos\eta\zhat$ are the radial and azimuthal coordinate vectors, respectively. Later we will use the fact that $\zehat= \bbeta+\Oh(u_s)$, $\bnu=\rhat+\Oh(u_s)$, and  $\shat=\xhat+\Oh(u_s)$, i.e.~recovering cylindrical coordinates at leading-order. 

\subsection{Force balance and filament dynamics}

Force balance is considered at each segment along the filament. At a station $s$, in the limit of large viscous damping relative to the inertial effects (i.e.,~Stokes flow), the resultant force  (per unit length) acting on the filament is zero at all times. We decompose this force as
\begin{gather}\label{eq:forcebalance}
  \f(s,t)=  \f^{\text{viscous}}(s,t)+\partial_s\F(s,t) + \f^{\text{LC}}(s,t)=\b{0},
\end{gather}
where  $\f^{\text{viscous}}$ is the viscous drag, $\partial_s\bm{F}$ is the differential elastic force, and $\f^{\text{LC}}$ is the LC restoring force. We shall compute each of these terms in turn.

First is the viscous drag, which we model using a resistive force approximation, 
\begin{gather}
    \f^{\text{viscous}} = -\bm{\mathcal{R}}\cdot \V,
\end{gather}
where $\bm{\mathcal{R}}$ is a resistance tensor,
\begin{gather}    
\bm{\mathcal{R}}=\zeta_{\parallel}\shat\shat+\zeta_{\perp}(\I-\shat\shat).
\end{gather}
The drag coefficients $\zeta_{\parallel}$ and $\zeta_{\perp}$ penalize motion in the filament tangent and normal directions, respectively. In terms of the rigid body motion and growth components of the filament velocity, we may write
\begin{gather}
\f^{\text{viscous}} = -\left[\zeta_{\parallel}\shat\shat+\zeta_{\perp}\left(\I-\shat\shat\right)\right]\cdot \left(\v+\alpha s \shat\right)
=-\zeta_{\parallel}(v_{\parallel}+\alpha s) \shat-\zeta_{\perp}\v_\perp.
\end{gather}
For reference, in a Newtonian fluid with viscosity $\mu$, we would have $\zeta_{\parallel} \approx 2\pi \mu/\log(L/2a)$ and $\zeta_{\perp}\approx 2\zeta_{\parallel}$ at leading-order in $1/\log(L/2a)$, assumed small; or,  we  instead have
$\zeta_{\parallel} \approx 2\pi \mu/\log(2h/a)$ and  $\zeta_{\perp}\approx 2\zeta_{\parallel}$ if the filament is moving a distance $h\ll L$ away from a wall \cite{wigg19}.

Next is the differential elastic force due to the connectivity of the filament itself. We assume that the filament cannot support a bending moment; hence, from the angular momentum balance, we can write the force density as a tangential piece alone, $\F(s,t) = T(s,t)\shat$.  With this, we can decompose the force balance, Eq.~\eqref{eq:forcebalance}, into its tangential and normal components,
\begin{subequations}
\begin{gather}
    -\zeta_{\parallel}(v_{\parallel}+\alpha s) +\partial_s T=0,\label{eq: force_balance_tangential}\\
    -\zeta_{\perp}\v_{\perp} + (\I-\shat\shat) \cdot \left(\b{f}^{\text{LC}}+T\partial_s\shat\right)=\bm{0},\label{eq:Force balance vertical}
\end{gather}
\end{subequations}
respectively, where we have used that $ \shat \cdot \b{f}^{\text{LC}}=0$ (see \S\ref{sec:LCTraction}). If both ends of the filament are free, the  end tensions  must equate, $T(\pm L/2)=T_{L}=\Oh(w^2)$, with $w=aW/K$ the dimensionless anchoring strength. Integrating the tangential force balance, Eq.~\eqref{eq: force_balance_tangential},  thus yields $v_{\parallel}=0$, as well as an expression for the tension,
\begin{gather}\label{eq:Tension}
    T(s,t) = T_{L}-\frac{\zeta_{\parallel}\alpha}{2}\left(\frac{L(t)^2}{4}-s^2\right).
\end{gather}
This baseline tension can be negative (compression), which has the potential to buckle the filament. This compression can be even more pronounced if one end of the filament meets additional resistance, due to surface interactions, for instance.

Consider the extreme case when one endpoint (e.g.,~the one at $s=-L/2$) is pinned (hinged), which requires an additional tangential force $F$ there. With $\V(-L/2,t)=[v_{\parallel}(t)-\alpha L(t)/2]\xhat+\v^\perp(-L/2,t)=\b{0}$, we find that $v_{\parallel}(t)=\alpha L/2$ and $\v^\perp(-L/2,t)=\b{0}$. Integrating \eqref{eq: force_balance_tangential}, we find that $   -\zeta_{\parallel}\alpha L^2/2+T_{L}-T_{-L}=0$  for $T_{\pm L}\equiv T(\pm L/2)$. Using the endpoint force balance, $F+T_{-L}=0$, we find that the required tangential force is $T_{-L}=-F=T_{L}-\zeta_{\parallel}\alpha L^2/4$. In this pinned problem the baseline tension is  given instead by
\begin{gather}\label{eq:Tension_pinned}
    T(s,t) =T_{L}-\frac{\alpha  \zeta_{\parallel}}{2}\left(\frac{L(t)}{2}-s\right) \left(\frac{3 L(t)}{2}+s\right).
\end{gather}
For the free filament, from \eqref{eq:Tension}, the tension is smallest (and becomes negative, as a compression) at the midpoint, $s=0$. For the pinned filament, from \eqref{eq:Tension_pinned} the tension is smallest at the point of pinning, $s=-L/2$, and it can be up to 4 times larger in the absence of  $T_{L}$ (i.e.~$-\alpha\zeta L^2/8$ vs.\ $-\alpha\zeta L^2/2$).

In either case, using $\v_\perp = u_t \yhat[1+\Oh(u_s)]$,  Eq.~\eqref{eq:Force balance vertical} may be written  as
\begin{gather}
\zeta_{\perp}u_t =T(s)u_{ss}+\yhat \cdot \b{f}^{\text{LC}}(s),\label{eq: filament_dynamics}
\end{gather}
where $T(s)$ is given in Eq.~\eqref{eq:Tension}, for free-free ends, or Eq.~\eqref{eq:Tension_pinned}, for fixed-free ends. In the free-free case, the   boundary conditions are selected to match the LC director field far away (assumed to be  the $\xhat$-direction), i.e.,~$u_s(\pm L/2)=0$. In the fixed-free case, the boundary conditions instead take the form $u(-L/2)=u_s(L/2)=0$.

\subsection{LC director field and traction}\label{sec:LCTraction}

Before we can determine the traction on the filament due to the surrounding LC environment, we need to determine the LC configuration. Fortunately, the system is evolving slowly, so  we can determine the LC field in a quasi-static manner, i.e.,~solving an equilibrium problem for each filament configuration. Here, we let $L=L(t)$ and drop all arguments of time since the LC director field is determined instantaneously at any given time $t$.

Far from the bacterial chain, the LC director field  is assumed to be uniform in the $\xhat$ direction. Elsewhere, at a position $\x$ in the fluid, the LC director field is represented as $\n(\x)= \xhat+\b{p}(\x)+\Oh(u_s^2)$ with $\b{p}\cdot\xhat=\bm{0}$, where $\b{p}=\Oh(u_s)$ in order to have a  balance between the deflection and the director field. With this approximation, the Ericksen torque stress tensor and projected molecular field are given by
\begin{subequations}
\begin{gather}
    \bPi=K\nabla\bm{n}=K\nabla\b{p}\left[1+\Oh(u_s)\right],\\
    \bm{h}=\P(\n)\cdot(\nabla\cdot\bPi)=K\nabla^2\b{p}\left[1+\Oh(u_s)\right],
\end{gather}
\end{subequations}
respectively, for the single (Frank) elastic constant $K$ and projection operator $\P(\n)\equiv\I-\n\n$, see Ref.~\cite{dp93}. At equilibrium, $\h=\bm{0}$ and so $\nabla^2\b{p}=\bm{0}$ with $\b{p}\to\bm{0}$ as $|\x|\to\infty$. On the boundary, $\bm{x}\in\partial D$, we have  the projected surface molecular field
\begin{equation}\label{eq:bcnematic}
    \bm{h}_{\text{surf}}=\P(\n)\cdot\left[\bnu\cdot\bPi-\frac{\partial\mathcal{F}_{\text{surf}}}{\partial\n}\right]=\left[K\partial_\nu\b{p}-\P(\n)\cdot\frac{\partial\mathcal{F}_{\text{surf}}}{\partial\n}\right]\left[1+\Oh(u_s)\right],
\end{equation}
where $\mathcal{F}_{\text{surf}}$ is a surface  energy that describes how the director field is anchored to the bacteria (to be defined), see Ref.~\cite{cs24c}. At equilibrium, $\bm{h}_{\text{surf}}(\bm{x})=\bm{0}$ for $\bm{x}\in \partial D$. Note from Eq.~\eqref{eq:xpos_asym} that  $\bm{h}_{\text{surf}}(\bm{x})=\bm{h}_{\text{surf}}(s\xhat+a\rhat)+\Oh(u\nabla\bm{h}_s)$  for $\x\in \partial D$, so $\partial D$ is effectively a  cylinder of radius $a$ aligned with $\xhat$.  We thus introduce a cylindrical polar coordinate system $(s,r,\eta)\in[-L/2,L/2]\times[a,\infty)\times[0,2\pi)$.

\subsubsection{Surface anchoring and director field}

On the filament surface, the LC is assumed to satisfy weak (i.e.~finite strength) degenerate tangential anchoring conditions, with surface energy
\begin{equation}
    \mathcal{F}_{\text{surf}}=\frac{W}{2}(\n\cdot\bnu)^2,
\end{equation}
for the unit normal vector $\bnu=\rhat-u_s\cos\eta\xhat +\Oh(u_s^2)$ and anchoring strength $W$, see Ref.~\cite{Rapini1969}. Since $\n\cdot\bnu=\b{p}\cdot\rhat-u_s\cos\eta+\Oh(u_s^2)$,  the surface molecular field is  
\begin{equation}\label{eq:degenbc}
    \bm{h}_{\text{surf}}=\left[K\partial_r\b{p}-W(\b{p}\cdot\rhat-u_s\cos\eta)\rhat\right]\left[1+\Oh(u_s)\right].
\end{equation}
Furthermore, since $\b{p}\cdot\xhat=0$, we can write $\b{p}=\phi_1(\x)\yhat+\phi_2(\x)\zhat$, for two scalar functions $\phi_1=\Oh(u_s)$ and $\phi_2=\Oh(u_s)$, which satisfy $\nabla^2\phi_1=\nabla^2\phi_2=0$ subject to $\phi_1(\x),\phi_2(\x)\to0$ as $|\x|\to\infty$ and
\begin{subequations}
\begin{gather}
  K\partial_r\phi_1=W\left[(\phi_1-u_s)\cos\eta+\phi_2\sin\eta\right]\cos\eta,\\
  K\partial_r\phi_2=W\left[(\phi_1-u_s)\cos\eta+\phi_2\sin\eta\right]\sin\eta,
\end{gather}
\end{subequations}
on $r=a$.

\paragraph{Free-free case:} In the case of two free-filament ends, with $u_s(\pm L/2)=0$, we can write $u$ as a cosine series:
\begin{gather}\label{eq:useries}
    u(s,t)=\sum_{k=1}^\infty \hat{u}_k(t) \cos\left(\frac{q_k(2s+L)}{L}\right),
\end{gather} 
where $q_k\equiv k\pi/2$. Using separation of variables, we pose general series representations for $\phi_1$ and $\phi_2$ of the form:
\begin{subequations}
\begin{align}
    \phi_1(\bm{x})&=\sum_{k=1}^\infty\frac{-2q_k\hat{u}_k}{L}\sin\left(\frac{q_k (2s+L)}{L}\right)\left[\sum_{m=-\infty}^\infty \frac{K_m(2q_k r/L)}{K_m(2q_k a/L)}a_k^me^{i m \eta}\right] ,\\
    \phi_2(\bm{x})&=\sum_{k=1}^\infty \frac{-2q_k\hat{u}_k}{L}\sin\left(\frac{q_k (2s+L)}{L}\right)\left[\sum_{m=-\infty}^\infty\frac{K_m(2q_k r/L)}{K_m(2q_k a/L)}b_k^me^{i m \eta}\right],
\end{align}
\end{subequations}
where $K_m(\cdot)$ is the $m$th modified Bessel function of the second kind and $a_k^m=\overline{a_k^{-m}}$ and $b_k^m=\overline{b_k^{-m}}$ are complex constants, which are to be determined now. 

The boundary conditions on $r=a$ impose the conditions
\begin{subequations}
\begin{gather}
    \sum_{m=-\infty}^\infty e^{i m \eta}\left[\frac{S_k^m}{w}a_k^m+\left(a_k^m\cos\eta+b_k^m\sin\eta\right)\cos\eta\right]=\cos^2\eta,\\
    \sum_{m=-\infty}^\infty e^{i m \eta}\left[\frac{S_k^m}{w}b_k^m+\left(a_k^m\cos\eta+b_k^m\sin\eta\right)\sin\eta\right]=\cos\eta\sin\eta,
\end{gather}
\end{subequations}
where $w=aW/K$ is the dimensionless anchoring strength, $S_k^m=S_{m}(2q_k a /L)$, and
\begin{equation}
    S_m(x)\equiv-\frac{x K_m'(x)}{K_m(x)}.
\end{equation}
Replacing the trigonometric functions with exponentials and evaluating coefficients yields the conditions
\begin{subequations}
\begin{gather}
  2a_k^m\left[1+\frac{2}{w}S_k^m\right]+a_k^{m-2}+a_k^{m+2}-\im b_k^{m-2}+\im b_k^{m+2}=2\delta_{0m}+\delta_{2m}+\delta_{-2m },\\
    2\im b_k^m\left[1+\frac{2}{w}S_k^m\right]+ a_k^{m-2} - a_k^{m+2} -\im b_k^{m-2}-\im b_k^{m+2}= \delta_{2m}-\delta_{-2m}.
\end{gather}
\end{subequations}
 where $\delta_{ij}$ is the Kronecker delta function. 
Solving these recurrence relations yields  $a_k^m=b_k^m=0$ for $m\notin\{-2,0, 2\}$, $b_k^0=0$, $\im b_k^2=\im b_k^{-2}= a_k^{-2}=a_k^{2}$,
\begin{gather}
    a_k^{0}=\frac{S_k^{2}}{S_k^{0}+S_k^{2}+\frac{2}{w}S_k^{0}S_k^{2}},\quad\text{and}\quad
 a_k^{2}=\frac{S_k^{0}/2}{S_k^{0}+S_k^{2}+\frac{2}{w}S_k^{0}S_k^{2}}.
\end{gather}
Altogether, we have  the solution
\begin{subequations}\label{eq:degsol}
\begin{align}
    \phi_1(\bm{x})&=\sum_{k=1}^\infty \frac{-2q_k\hat{u}_k}{L}\sin\left(\frac{q_k (2s+L)}{L}\right)\left[A_k\frac{K_0(2q_k r/L)}{K_0(2q_k a/L)}+B_k\cos(2\eta)\frac{K_2(2q_k r/L)}{K_2(2q_k a/L)} \right],\\
    \phi_2(\bm{x})&=\sum_{k=1}^\infty \frac{-2q_k\hat{u}_k}{L}\sin\left(\frac{q_k (2s+L)}{L}\right)\left[B_k\sin(2\eta)\frac{K_2(2q_k r/L)}{K_2(2q_ka/L)} \right],
\end{align}
\end{subequations}
where 
\begin{equation}\label{eq:Akfreefree}
    A_k\equiv\frac{1/S_0(2q_ka/L)}{1/S_2(2q_ka/L)+1/S_0(2q_ka/L)+2/w}\quad\text{and}\quad  B_k\equiv\frac{1/S_2(2q_ka/L)}{1/S_2(2q_ka/L)+1/S_0(2q_ka/L)+2/w}.
\end{equation}

\paragraph{One free-end:} In the case of one fixed and one free-filament ends, with $u(-L/2)=0$ and $u_s(L/2)=0$, respectively, we can instead write $u$ as the series
\begin{gather}
    u(s,t)=\sum_{k=1}^\infty \hat{u}_k(t) \sin\left(
    \frac{Q_k(2s+L)}{L}\right),
\end{gather} 
for $Q_k\equiv (2k-1)\pi/4$. In this case, we find
\begin{subequations}\label{eq:degsol_onefree}
\begin{align}
    \phi_1(\bm{x})&=\sum_{k=1}^\infty \frac{2Q_k\hat{u}_k}{L}\cos\left(\frac{Q_k (2s+L)}{L}\right)\left[C_k\frac{K_0(2Q_k r/L)}{K_0(2Q_k a/L)}+D_k\cos(2\eta)\frac{K_2(2Q_k r/L)}{K_2(2Q_k a/L)} \right],\\
    \phi_2(\bm{x})&=\sum_{k=1}^\infty \frac{2Q_k\hat{u}_k}{L}\cos\left(\frac{Q_k (2s+L)}{L}\right)\left[D_k\sin(2\eta)\frac{K_2(2Q_k r/L)}{K_2(2Q_ka/L)} \right],
\end{align}
\end{subequations}
where 
\begin{equation}
    C_k\equiv\frac{1/S_0(Q_k2a/L)}{1/S_2(2Q_ka/L)+1/S_0(2Q_ka/L)+2/w}\quad\text{and}\quad  D _k\equiv\frac{1/S_2(2Q_ka/L)}{1/S_2(2Q_ka/L)+1/S_0(2Q_ka/L)+2/w}.
\end{equation}

\subsubsection{Surface traction}
To compute the traction on the filament due to the LC, we will need the Ericksen stress tensor and the surface stress tensor, given by 
\begin{subequations}
\begin{gather}
\Tm = \frac{K}{2}\lVert\nabla\n\rVert^2\mathbf{I}-K\nabla\n\cdot\nabla\n^T,\label{eq:EricksenStress}\\
\Tm_{\mathrm{surf}} = \mathcal{F}_{\mathrm{surf}} \P(\bnu)-\P(\bnu)\cdot\frac{\partial \mathcal{F}_{\mathrm{surf}}}{\partial\bnu}\bnu,
\end{gather}
\end{subequations}
respectively, see \cite{dp93,cs24c}.

The first part of the traction is associated with the Ericksen stress $\Tm$, and is given by $\f(s,\eta) = \bnu\cdot \Tm$. But this traction enters at $\Oh(\b{p}^2)=\Oh(u_s^2)$, and so we neglect it. The second part of the traction is associated with the surface stress $\Tm_{\mathrm{surf}}$, and is given by $\f_{\mathrm{surf}}(s,\eta) = \nabla_s \cdot  \Tm_{\mathrm{surf}}$,
where $\nabla_s \equiv \P(\bnu)\cdot \nabla$ with $\P(\bnu)\equiv\I-\bnu\bnu$, a projection operator for the surface normal direction.  Since $\bm{\bnu}=\rhat+\Oh(u_s)$, we  note that
\begin{equation}\label{eq:nablas}
    \nabla_s=\P(\bnu)\cdot\nabla=\left(\xhat\partial_s+\frac{1}{r}\bbeta\partial_\eta\right)\left[1+\Oh(u_s)\right].
\end{equation}

For the weak degenerate tangential anchoring conditions  considered above, we have the surface stress 
\begin{equation}
    \Tm_{\mathrm{surf}}=\frac{W}{2}(\n\cdot\bnu)^2\P(\bnu)-W(\n\cdot\bnu)\P(\bnu)\cdot\n\bnu,
\end{equation}
for $\bnu=\rhat-u_s\cos\eta\xhat +\Oh(u_s^2)$ and $\n=\xhat+\phi_1\yhat+\phi_2\zhat+\Oh(u_s^2)$. It follows that $\n\cdot\bnu=(\phi_1-u_s)\cos\eta+\phi_2\sin\eta+\Oh(u_s^2)$, and so,
\begin{equation}
    \Tm_{\mathrm{surf}}=-W[(\phi_1-u_s)\cos\eta+\phi_2\sin\eta]\xhat\rhat\left[1+\Oh(u_s)\right]=-K\partial_r\left[\phi_1\xhat\yhat+\phi_2\xhat\zhat\right]\left[1+\Oh(u_s)\right],
\end{equation}
where we have used the boundary condition \eqref{eq:degenbc}. Using \eqref{eq:nablas} yields the surface traction:
\begin{equation}
   \f_{\mathrm{surf}}(s,\eta)=\nabla_s\cdot\Tm_{\mathrm{surf}}=-K\left[\partial_s\partial_r\phi_1\yhat+\partial_s\partial_r\phi_2\zhat\right]\left[1+\Oh(u_s)\right].
\end{equation}

Finally, the force (per length) acting on the filament at each station $s$, i.e.~$\f^{\mathrm{LC}}(s)$, is found by integrating the total surface traction, $\f(s,\eta)+\f_{\mathrm{surf}}(s,\eta)$, over the azimuthal angle, $\eta$:
\begin{equation}
    \f^{\mathrm{LC}}(s)=\int_0^{2\pi} \left[\f(s,\eta)+\f_{\mathrm{surf}}(s,\eta)\right]a\de\eta=-aK\left[\left(\int_0^{2\pi}\partial_s\partial_r\phi_1 \de \eta\right)\yhat+\left(\int_0^{2\pi}\partial_s\partial_r\phi_2 \de \eta\right)\zhat\right]\left[1+\Oh(u_s)\right].
\end{equation}
Note that $\shat\cdot \f^{\mathrm{LC}}(s)=\Oh(u_s^2)$, as used earlier. 

\paragraph{Two free ends:}  For two free ends, we have from \eqref{eq:degsol},
\begin{subequations}
\begin{align}
    \partial_s\partial_r\phi_1&=\sum_{k=1}^\infty \left(\frac{2q_k}{L}\right)^2\frac{\hat{u}_k}{a}\cos\left(\frac{q_k (2s+L)}{L}\right)\left[A_kS_0\left(\frac{2q_k a}{L}\right)+B_k\cos(2\eta)S_2\left(\frac{2q_k a}{L}\right) \right],\\
   \partial_s\partial_r\phi_2&=\sum_{k=1}^\infty  \left(\frac{2q_k}{L}\right)^2\frac{\hat{u}_k}{a}\cos\left(\frac{q_k(2s+L)}{L}\right)\Big[B_k\sin(2\eta)S_2\left(\frac{2q_k a}{L}\right) \Big],
\end{align}
\end{subequations}
on $r=a$, for $q_k=\pi k /2$. Using that $\cos(2\eta)$ and $\sin(2\eta)$ have zero mean, and inserting the expressions in \eqref{eq:Akfreefree}, the LC restoring force is found to be
\begin{gather}\label{eq:f_LC_degenerate}
   \f^{\text{LC}}(s)\sim -\frac{16 \pi a K}{L^3}\yhat\sum_{k=1}^\infty q_k^3\hat{u}_k\left[\frac{4q_k a}{wL}-\frac{K_0(2q_k a/L)}{K_0'(2q_k a/L)}-\frac{K_2(2q_k a/L)}{K_2'(2q_k a/L)}\right]^{-1}\cos\left(\frac{q_k (2s+L)}{L}\right).
\end{gather}

\paragraph{Fixed-free ends:} For only one free-end, the restoring force is similarly found to be  \eqref{eq:f_LC_degenerate}, but with the substitutions $\cos \mapsto \sin$ and $q_k \mapsto Q_k=(2k-1)\pi/4$.

\subsection{Filament dynamics and buckling length estimates}

The filament dynamics in Eq.~\eqref{eq: filament_dynamics} are now fully determined, and we can begin to analyze them. Let us at this stage non-dimensionalize the problem by defining a dimensionless time, $\tau\equiv \alpha t$, dimensionless length, $\xi\equiv 2s/L(t)$, where $\xi\in[-1,1]$, and  dimensionless displacement  $U(\xi,\tau)\equiv u(s,t)/a$.  We also introduce the dimensionless filament length $\Lambda(\tau)\equiv L(t)/(2a)$, dimensionless viscosity $\chi\equiv \zeta_\perp/\zeta_\parallel$,   dimensionless  anchoring strength $w\equiv aW/K$, and Ericksen number $\Er\equiv a^2\zeta_\parallel \alpha/(2\pi K)$.

From the uniform growth law, $L'(t)=\alpha L(t)$ (see \S\ref{sec:kinematics}), we  have that
\begin{equation}
    \partial_t u(s,t) =a\alpha\left[\partial_\tau U(\xi,\tau)-\xi \partial_\xi U(\xi,\tau) \right],
\end{equation}
and so the filament dynamics, i.e.,~Eq.~\eqref{eq: filament_dynamics}, is governed by the dimensionless equation
\begin{equation}\label{eq:dimensionlessDynamics}
    \chi(U_\tau-\xi  U_\xi) =\tilde{T}(\xi)U_{\xi\xi}+ \Er^{-1} F^{\mathrm{LC}}(\xi,\tau),
\end{equation}
for the dimensionless tension
\begin{equation}
\tilde{T}(\xi)\equiv \frac{T(s,t)}{\alpha \zeta_\parallel L(t)^2/4}=
    \begin{cases}
      -\left(1-\xi^2\right)/2 &\text{free-free,}\\
       -\left(1-\xi\right)\left(3+\xi\right)/2 &\text{fixed-free,}
    \end{cases}
\end{equation}
where we have dropped the end-force $T_L$ as lower order, the dimensionless LC restoring force 
\begin{gather}
F^{\mathrm{LC}}(\xi,\tau)\equiv \frac{f^{\mathrm{LC}}(s,t)}{2\pi K/a} =
\begin{dcases}
-\sum_{k=1}^\infty \frac{q_k^3 U_k}{\Lambda^3}\left[\frac{2q_k }{w\Lambda}-\frac{K_0(q_k/\Lambda)}{K_0'(q_k /\Lambda)}-\frac{K_2(q_k/\Lambda)}{K_2'(q_k /\Lambda)}\right]^{-1}\cos\left[q_k (1+\xi)\right]&\text{free-free,}\\
-\sum_{k=1}^\infty \frac{Q_k^3 U_k}{\Lambda^3}  \left[\frac{2Q_k }{w\Lambda}-\frac{K_0(Q_k/\Lambda)}{K_0'(Q_k /\Lambda)}-\frac{K_2(Q_k/\Lambda)}{K_2'(Q_k /\Lambda)}\right]^{-1}\sin\left[Q_k(1+\xi)\right] &\text{fixed-free,}
    \end{dcases}
\end{gather}
 with $q_k=k\pi/2$, $Q_k= (2k-1)\pi/4$, and the Fourier series
\begin{gather}
U(\xi,\tau) =
\begin{dcases}
\sum_{k=1}^\infty U_k(\tau)\cos\left[q_k (1
+\xi)\right]&\text{free-free,}\\
\sum_{k=1}^\infty U_k(\tau)\sin\left[Q_k (1+\xi)\right] &\text{fixed-free.}
    \end{dcases}
\end{gather}
with coefficients $U_k(\tau)\equiv u_k(t)/a$.

\subsubsection{Two free-ends}
In the free-free case, inserting the Fourier series for $U(\xi,\tau)$ and using the orthogonality of the basis functions yields
\begin{equation}\label{eq:dispersion}
  \chi  U_k'(\tau) =\left\{\frac{\chi}{2}+\frac{q_k^2}{3}-\frac{1}{4}-\frac{q_k^3}{\Er\Lambda^3}\left[\frac{2q_k }{w\Lambda}-\frac{K_0(q_k/\Lambda)}{K_0'(q_k /\Lambda)}-\frac{K_2(q_k/\Lambda)}{K_2'(q_k /\Lambda)}\right]^{-1}\right\}U_k(\tau)+\sum_{m\neq k}J_{km}U_m(\tau),
\end{equation}
for  $q_k=\pi k/2$ and
\begin{equation}
\begin{split}
    J_{km}&=\int_{-1}^{1}\left\{\frac{q_m^2}{2}(1-\xi^2)\cos\left[q_m (1+\xi)\right]-\chi q_m \xi \sin\left[q_m (1+\xi)\right]\right\}\cos\left[q_k (1+\xi)\right]\de\xi\\
    &=-[1+(-1)^{k+m}]\frac{m^2[m^2(1-\chi)+k^2(1+\chi)]}{(k^2-m^2)^2},
    \end{split}
\end{equation}
when $m\neq k$. Here, we see that the even and odd modes about $\xi=0$ have uncoupled. Since there is fairly rapid decay in $k$ away from $k=m$, a first   approximation is to ignore the summation in  Eq.~\eqref{eq:dispersion}. Doing this yields
\begin{equation}
  \chi  U_k'(\tau) \approx\left\{\frac{\chi}{2}+\frac{q_k^2}{3}-\frac{1}{4}-\frac{q_k^3}{\Er\Lambda^3}\left[\frac{2q_k }{w\Lambda}-\frac{K_0(q_k/\Lambda)}{K_0'(q_k /\Lambda)}-\frac{K_2(q_k/\Lambda)}{K_2'(q_k /\Lambda)}\right]^{-1}\right\}U_k(\tau).
\end{equation}

Although $\Lambda(\tau)$ is generally time-dependent, to find the buckling lengths, we shall  freeze it at its instantaneous value, $\Lambda(\tau)= \Lambda$, and pose the ansatz $U_k(\tau)=e^{\sigma_k \tau}\hat{U}_k$, yielding the growth rate of the  $k$th mode:
\begin{equation}
  \sigma_k\approx\frac{1}{2}+\frac{1}{\chi}\left\{\frac{q_k^2}{3}-\frac{1}{4}-\frac{q_k^3}{\Er\Lambda^3}\left[\frac{2q_k }{w\Lambda}-\frac{K_0(q_k/\Lambda)}{K_0'(q_k /\Lambda)}-\frac{K_2(q_k/\Lambda)}{K_2'(q_k /\Lambda)}\right]^{-1}\right\}.
\end{equation}
It follows that the $k$th mode becomes unstable when $\Lambda>\Lambda_k$ where $\Lambda_k$ solves
\begin{equation}
    \frac{\chi}{2}+\frac{q_k^2}{3}-\frac{1}{4}-\frac{q_k^3}{\Er\Lambda_k^3}\left[\frac{2q_k }{w\Lambda_k}-\frac{K_0(q_k/\Lambda_k)}{K_0'(q_k /\Lambda_k)}-\frac{K_2(q_k/\Lambda_k)}{K_2'(q_k /\Lambda_k)}\right]^{-1}\approx 0.
\end{equation}

For small $\Er$ (and thus large $\Lambda_k$),  we  note that
\begin{equation}
  \frac{\Lambda_k^3}{q_k^3}\left[\frac{2 q_k}{w\Lambda_k}- \frac{K_0(q_k/\Lambda_k)}{K_0'(q_k /\Lambda)}-\frac{K_2(q_k/\Lambda_k)}{K_2'(q_k /\Lambda_k)}\right]=\frac{\Lambda_k^2}{2q_k^2}\left(1-2\gamma+\frac{4}{w}+\log\frac{4\Lambda_k^2}{q_k^2}\right)+\Oh\left(1\right),
\end{equation}
as $q_k/\Lambda_k\to 0$, where $\gamma$ is the Euler--Mascheroni constant.  At leading-order in $\Er^{-1}$, assumed large, we have that
\begin{equation}
  \frac{\Lambda_k^2}{2q_k^2}\left(1-2\gamma+\frac{4}{w}+\log\frac{4\Lambda_k^2}{q_k^2}\right)\approx\frac{4\Er^{-1}}{2\chi-1+4q_k^2/3},
\end{equation}
which finally gives the approximation
\begin{equation}
\Lambda_k\approx\left(\frac{2\pi^2 k^2 \Er^{-1}}{2\chi-1+\pi^2k^2/3}\right)^{1/2}\left[W\left(\frac{32\Er^{-1}e^{1-2\gamma+4/w}}{2\chi-1+\pi^2k^2/3}\right)\right]^{-1/2},
\end{equation}
where $W(z)$ is the Lambert $W$ function (the solution to $W\exp W=z$). Using the rudimentary approximation of $W(z)\approx \log z$ for large $z$, this expression can be approximated further as
\begin{equation}\label{eq:decoupleapprox_freefree}
\Lambda_k\approx\left(\frac{2\pi^2 k^2 \Er^{-1}}{2\chi-1+\pi^2k^2/3}\right)^{1/2}\left[\log\left(\frac{32\Er^{-1}}{2\chi-1+\pi^2k^2/3}\right)+1-2\gamma+\frac{4}{w}\right]^{-1/2}.
\end{equation}

The buckling length of the filament is then given by the smallest  critical length,  $\Lambda^*=\min_k \Lambda_k$. For $\chi=5$, $\Er=10^{-6}$, and   $w=10^{-2}$--$10^0$, we find that the smallest mode is $k=1$, which gives the estimate to dimensionless buckling $\Lambda^*\approx62.2$--$294$. This approximation is compared to numerical solutions of the coupled system, Eq.~\eqref{eq:dispersion}, in the main text Fig.~5. Although the numerical results are not quantitatively  recovered, the  approximation  does recover the  qualitative behavior. It thus inspires  the fitting expression presented in the main text:
\begin{equation}\label{eq:fitting}
    \Lambda^* = \left(\frac{a_1 \Er^{-1}}{\log( a_2 \Er^{-1})+4/w}\right)^{1/2},
\end{equation}
for fitting parameters $a_i$.

\subsubsection{Fixed-free ends}

In the fixed-free case, inserting the Fourier series for $U(\xi,\tau)$ and using the orthogonality of the basis functions yields a similar expression to before:
\begin{equation}\label{eq:dispersion2}
  \chi  U_k'(\tau) =\left\{\frac{4Q_k^2}{3}-\frac{1}{4}-\frac{Q_k^3}{\Er\Lambda^3}\left[\frac{2Q_k }{w\Lambda}-\frac{K_0(Q_k/\Lambda)}{K_0'(Q_k /\Lambda)}-\frac{K_2(Q_k/\Lambda)}{K_2'(Q_k /\Lambda)}\right]^{-1}\right\}U_k(\tau)+\sum_{m\neq k}I_{km}U_m(\tau),
\end{equation}
for  $Q_k=\pi (2k-1)/4$ and
\begin{equation}
\begin{split}
    I_{km}&=\int_{-1}^{1}\left\{\frac{Q_m^2}{2}(1-\xi)(3+\xi)\sin\left[Q_m (1+\xi)\right]+\chi Q_m \xi \cos\left[Q_m (1+\xi)\right]\right\}\sin\left[Q_k (1+\xi)\right]\de\xi\\
    &=-\frac{(-1)^{k+m}(2m-1)^2\left[(2k-1)^2+(2m-1)^2\right]}{8(k-m)^2(k+m-1)^2}-\frac{(2m-1)\left[2k-1+(-1)^{k+m}(2m-1)\right]\chi}{4(k-m)(k+m-1)},
    \end{split}
\end{equation}
when $m\neq k$.

Again, omitting the summation in Eq.~\eqref{eq:dispersion2} and taking the ansatz $U_k(\tau)=e^{\sigma_k \tau}\hat{U}_k$ yields an approximation for the growth rate of the $k$th mode,
\begin{equation}
    \sigma_k\approx\frac{1}{\chi}\left\{\frac{4 Q_k^2}{3}-\frac{1}{4}- \frac{ Q_k^3}{\Er\Lambda^3}\left[\frac{2Q_k }{w\Lambda}-\frac{K_0(Q_k/\Lambda)}{K_0'(Q_k /\Lambda)}-\frac{K_2(Q_k/\Lambda)}{K_2'(Q_k /\Lambda)}\right]^{-1}\right\}.
\end{equation}

For small $\Er$, making the same approximations as  before yields the fact that the $k$th mode becomes unstable when $\Lambda>\Lambda_k$, where
\begin{equation}
\Lambda_k\approx\left(\frac{(2k-1)^2 \pi^2\Er^{-1}/2}{\pi^2(2k-1)^2/3 -1}\right)^{1/2}\left[W\left(\frac{32 \Er^{-1}}{\pi^2(2k-1)^2/3-1} e^{1-2\gamma+4/w}\right)\right]^{-1/2},
\end{equation}
or, with the additional rudimentary approximation,
\begin{equation}
\Lambda_k\approx\left(\frac{(2k-1)^2 \pi^2\Er^{-1}/2}{\pi^2(2k-1)^2/3 -1}\right)^{1/2}\left[\log\left(\frac{32 \Er^{-1}}{\pi^2(2k-1)^2/3-1} \right)+1-2\gamma+\frac{4}{w}\right]^{-1/2}.
\end{equation}

For $\chi=5$, $\Er=10^{-6}$, and   $w=10^{-2}$--$10^0$, here the smallest value of $\Lambda_k$ is found to range between  when $k=6$ (for $w=10^{-2}$) and $k=2$   (for $w=10^{0}$), giving the estimate to dimensionless buckling $\Lambda^*\approx60.5$--$296$. An improved estimation is found by fitting \eqref{eq:fitting} to the coupled system, as  presented in the main text.

\begin{figure*}
    \centering
    \includegraphics[width=.9\linewidth]{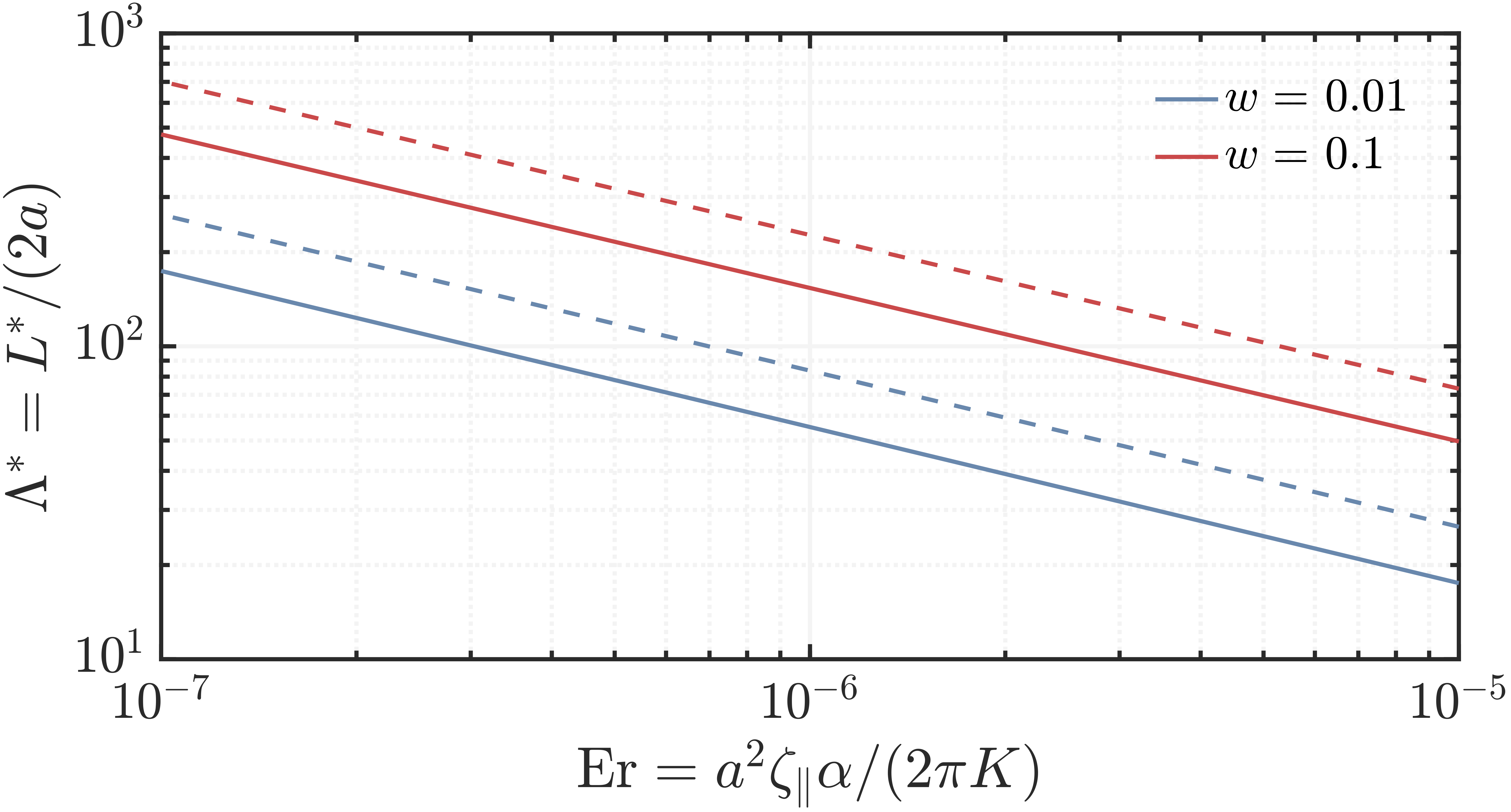}
    \caption{ {The analytical approximation [Eq.~\ref{eq:fitting}] for the dimensionless buckling length, $\Lambda^*$, as a function of the Ericksen number, $\Er$, for free-free (dashed curves) and fixed-free (solid curves) boundary conditions. A fixed end reduces the buckling length by a factor of $\approx1.5$.}}
    \label{fig:LstarComparison}
\end{figure*}

\clearpage

\section*{Supplementary movies}\label{simovies}
Supplementary movies are available on Zenodo (\url{https://doi.org/10.5281/zenodo.17917861}).

\begin{enumerate}
    \item \textbf{Movie~S1:} Non-motile \textit{E.~coli} proliferating in LB media without DSCG.
    \label{vid:lb video growth}
    \item \textbf{Movie~S2:} Non-motile \textit{E.~coli} proliferating in LB media supplemented with \SI{15}{w/w\%} DSCG.
    \label{vid:dscg video growth}
    \item \textbf{Movie~S3:} Non-motile \textit{E.~coli} proliferating in LB media supplemented with \SI{15}{w/w\%} DSCG. At $t = 0$~min we flush a suspension of chains with LC free liquid at a flow rate of \SI{2.5}{\uL/\min}, thus decreasing the LC concentration as time increases.
    \label{vid:dilution video}
    \item \textbf{Movie~S4:} Non-motile \textit{E.~coli} proliferating in LB media supplemented with \SI{15}{w/w\%} DSCG. At $t = 0$~min we flush a suspension of chains with
    \SI{15}{w/w\%} DSCG solution at a flow rate of~\SI{2.5}{\uL/\min} keeping the LC concentration constant as time increases.
\end{enumerate}

\clearpage
\def\arraystretch{1.5}
\begin{table}[p]
    \begin{center}
    \begin{tabular}{|c|c|c|}
    \hline
         \textbf{Parameter description} & \textbf{Symbol} & \textbf{Value} \\
         \hline
         Cell growth rate & $\alpha$ & \SI{2e-4}{\sec^{-1}}\\
         \hline
         DSCG parallel viscosity & $\mu_\parallel$ & \makecell{\SI{0.4}{\Pa\cdot\sec}~\cite{habibi2019passive} (\SI{18}{w/w\%}) \\
         \SI{0.1}{\Pa\cdot\sec}~\cite{habibi2019passive} (\SI{15}{w/w\%})} \\
         \hline
         {DSCG viscosity ratio} & {$\mu_\perp/\mu_\parallel$} & {$2$--$6$ (measured, cf.~\cite{duchesne2015bacterial,Gomez2016})}\\ 
         \hline
         Cell radius & $a$ & \SI{1}{\um}  \\
         \hline
         LC - bacteria surface anchoring energy & $W$ & \makecell{\SI{e-5}{\J/\m^2}~\cite{mushenheim_dynamic_2014} \\
         \SI{e-6}{\J/\m^2}~\cite{zhou2017dynamic}\\
         \SI{0.2e-5}{\J/\m^2}~\cite{chi2020surface}}\\
         \hline
         DSCG average Frank elastic constant & $K$ & \makecell{\SI{18.66}{\pico\newton}~\cite{zhou2014living} (\SI{18}{w/w\%}) \\
         \SI{11.91}{\pico\newton}~\cite{zhou2014living} (\SI{15}{w/w\%})}  \\
         \hline
         Volume fraction & $\phi$ & 0.115~\cite{zhou_elasticity_2014} \\
         \hline
         Average LC molecule diameter & $D$ & \SI{1.6}{\nm}~\cite{zhou_elasticity_2014} \\
         \hline
         Average LC molecule length & $l$ & \SI{20}{\nm}~\cite{zhou_elasticity_2014} \\
         \hline
         Cell length & $l_\mathrm{b}$ & $1$--\SI{4}{\um}  \\
         \hline
         Buckling length & $L^*$ & \makecell{\SI{87 \pm 42}{\um} (\SI{18}{w/w\%} DSCG)\\ 
         \SI{166\pm 93}{\um}(\SI{15}{w/w\%} DSCG)} \\
         \hline
         cell height from surface & $h$ & \makecell{\SI{1.0 \pm 0.4}{\um} (\SI{18}{w/w\%} DSCG)\\ 
         \SI{1.0 \pm 0.7}{\um} (\SI{15}{w/w\%} DSCG)} \\
        \hline
        \end{tabular}
        \label{table_expts}
        \caption{Summary of measurements and parameters used in theoretical calculations.}
        \label{tab: table_expts_th}
    \end{center}
\end{table}

\clearpage


\begin{figure*}
\includegraphics[width=0.3\textwidth]{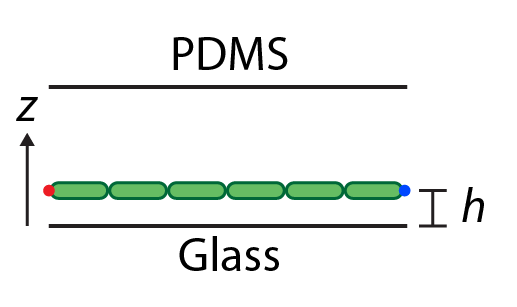}
\caption{\label{fig:h schematic}\textbf{Schematic depicting the geometry to estimate the distance between a bacterial chain and a surface.} The schematic shows a bacterial chain (green spherocylinders) within the custom PDMS device. The bacterial chain is a distance $h$ from the glass bottom surface. To estimate~$h$~we track the endpoints of the bacterial chain (indicated here as red and blue dots). Next, we take the standard deviation of the $z$-position time series to get an estimate of the chain's fluctuations near the surface of the imaging device. The average of the fluctuations yields an estimate of $h$ for a single chain. We repeat this process for multiple buckling chains to obtain the average values shown in Table~\ref{tab: table_expts_th}.}
\end{figure*}

\begin{figure*}
\includegraphics[width=\textwidth]{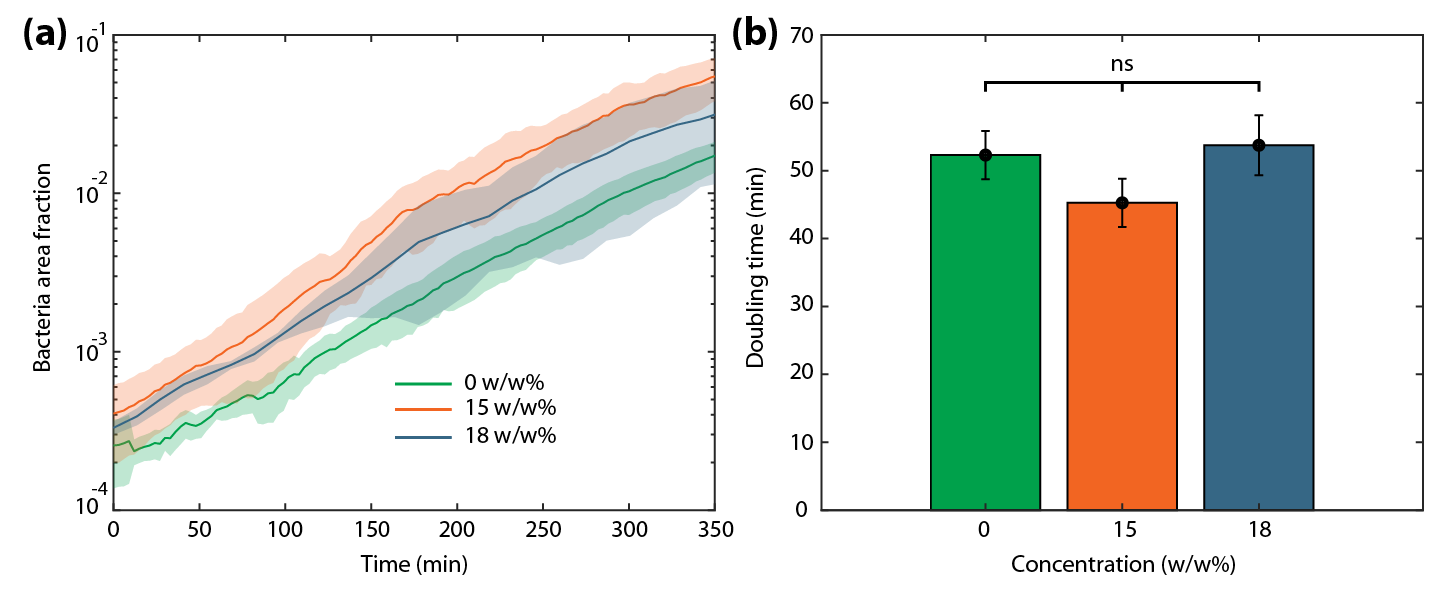}
\caption{\label{fig:growth_curves} \textbf{\textit{E.~coli} doubling times are similar across different LC conditions}. \textbf{(A)} and \textbf{(B)} show
the growth curves and doubling times for \textit{E.~coli} cells in different LC solutions. The growth curves were obtained in triplicate where the solid curve represents the mean among all
samples while the shaded regions are the corresponding standard deviations. The doubling times were measured by fitting the first 500 min of each sample’s growth curve to an exponential
equation: $A_\mathrm{frac} = A_0 e^{\alpha t}$ where $t$ is time, $A_0$ is the area fraction at $t = 0$, $\alpha$ is the maximal growth rate, and $A_\mathrm{frac}$ is the cell collective's total area fraction. We convert the bacteria’s maximal growth rate, $\alpha$, to a doubling time via $t_\mathrm{d} = \mathrm{log}(2)/\alpha$. We observe that, throughout
all LC concentrations tested, \textit{E.~coli}'s doubling time remains between 52--68 min and does not depend appreciably on LC concentration. We obtained the growth curves under
different LC concentrations via segmentation of images in MATLAB  to then plot the area fraction that cells occupy as a function of time.}
\end{figure*}

\begin{figure*}
\includegraphics[width=\textwidth]{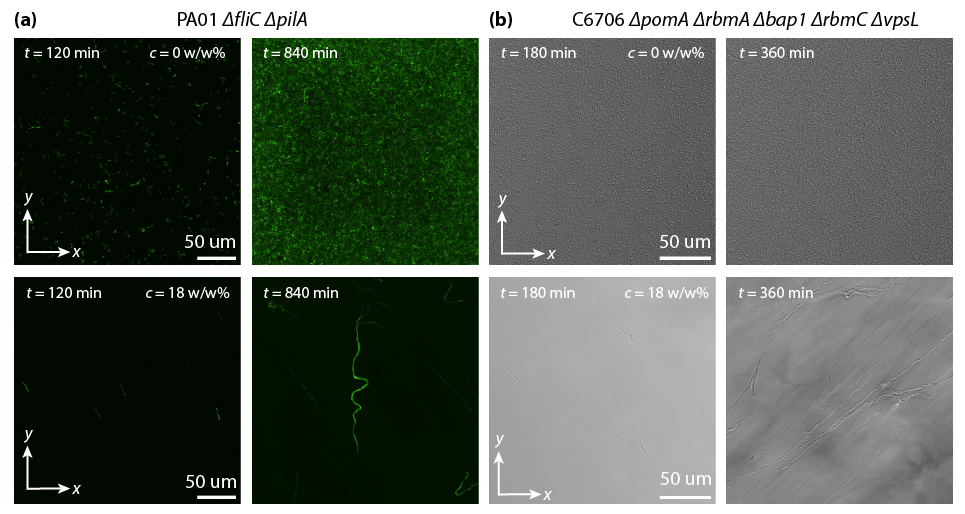}
\caption{\textbf{Nonmotile \textit{P.~aeruginosa} and \textit{V.~cholerae} form chains as they proliferate in liquid crystalline solutions.} \textbf{(A)} Time sequence of nonmotile \textit{P.~aeruginosa} proliferating in nutrient-rich fluid without (top) or with (bottom) \SI{18}{w/w\%} DSCG. \textbf{(B)} Same as in \textbf{A} but with nonmotile \textit{V.~cholerae}. The time indicates the duration elapsed after the 96-well plate containing the samples is inserted into the static incubator. The images were taken at different positions of the same well at the different time points.}
\label{fig:cholerae_pseudo_supp}
\end{figure*}

\clearpage

\providecommand{\noopsort}[1]{}\providecommand{\singleletter}[1]{#1}%

\end{document}